\documentstyle[11pt,epsf]{article}

\def\Split{\mathop{\rm Split}\nolimits}
\def\Soft{\mathop{\cal S}\nolimits}
\def\eps{\epsilon}
\def\tree{{\rm tree}}
\def\oneloop{{1 \mbox{-} \rm loop}}
\def\Li{\mathop{\hbox{\rm Li}}\nolimits}
\def\Ord{{\cal O}}

\def\pol{\varepsilon}
\def\qb{{\bar q}}
\def\A#1{{\cal A}_{#1}}
\def\si{\sigma}
\def\ib{{\bar\imath}}
\def\ns{n_{\mskip-2mu s}}\def\nf{n_{\mskip-2mu f}}
\def\fact{{\rm fact}}
\def\nonfact{{\rm non\hbox{-}fact}}
\def\Tr{\mathop{\rm Tr}\nolimits}

\def\spa#1.#2{\left\langle#1\,#2\right\rangle}
\def\spb#1.#2{\left[#1\,#2\right]}
\def\L{\left(}
\def\R{\right)}

\def\eqn#1{eq.~(\ref{#1})}
\def\eqns#1#2{eqs.~(\ref{#1}) and~(\ref{#2})}
\def\fig#1{fig.~{\ref{#1}}}
\def\sec#1{section~{\ref{#1}}}
\def\app#1{appendix~\ref{#1}}
\def\tab#1{table~\ref{#1}}

\def\feynsl#1{
  \setbox0=\hbox{/} \setbox1=\hbox{$#1$}
  \dimen0=\wd0 \advance\dimen0 by -\wd1 \divide\dimen0 by 2
  \ifdim\wd0>\wd1 \raise.15ex\copy0\kern-\wd0\kern\dimen0\copy1\kern\dimen0
  \else \kern-\dimen0\raise.15ex\copy0\kern-\dimen0\kern-\wd1\copy1\fi}
\def\dprod#1#2{{\left({#1}\!\cdot\!{#2}\right)}}
\def\iscol#1#2{{\buildrel#1\parallel#2\over\longrightarrow}}
\def\hypgeo{{\vphantom{F}}_2F_1}

\newskip\humongous \humongous=0pt plus 1000pt minus 100pt
\def\caja{\mathsurround=0pt}
\def\eqalign#1{\,\vcenter{\openup1\jot \caja
       \ialign{\strut \hfil$\displaystyle{##}$&$
        \displaystyle{{}##}$\hfil\crcr#1\crcr}}\,}
\newif\ifdtup

\makeatletter
\def\@eqnnum{\hbox{\reset@font\rm(\theequation)}}
\let\make@eqnnum=\@eqnnum %
\def\eqnum#1{\dec@eqnnum \global\def\make@eqnnum{\reset@font\rm(#1)}%
\def\@currentlabel{#1}%
}
\def\inc@eqnnum{\addtocounter{equation}{1}}
\def\dec@eqnnum{\addtocounter{equation}{-1}}
\@definecounter{equation}%
\@addtoreset{equation}{section} %
\def\theequation@prefix{{\thesection}.} %
\def\theequation{\theequation@prefix\arabic{equation}}%
\makeatother

\begin{document}
\begin{titlepage}

\noindent hep-ph/9903516
\hspace*{\fill}\parbox[t]{4cm}{
BNL-HET-99/6\\
DFTT 16/99\\
UCLA/99/TEP/11\\
MSUHEP-90325\\ 
\today}

\begin{center}
{\Large\bf The Infrared Behavior of One-Loop QCD Amplitudes at
Next-to-Next-to-Leading Order} \\
\vspace{1.cm}

{Zvi Bern}\\
\vspace{.2cm}
{\sl Department of Physics and Astronomy\\
University of California at Los Angeles\\
Los Angeles,  CA 90095-1547, USA}\\

\vspace{.8cm}
{Vittorio Del Duca}\\
\vspace{.2cm}
{\sl I.N.F.N., Sezione di Torino\\
via P. Giuria, 1-10125\\
Torino, Italy}\\

\vspace{.8cm}
{William B. Kilgore}\\
\vspace{.2cm}
{\sl Department of Physics\\
Brookhaven National Laboratory\\
Upton, NY 11973-5000, USA}\\

\vspace{.5cm}
and \\
\vspace{.5cm}

{Carl R. Schmidt}\\
\vspace{.2cm}
{\sl Department of Physics and Astronomy\\
Michigan State University\\
East Lansing, MI 48824, USA}\\

\vspace{.5cm}

\begin{abstract}
We present universal factorization formulas describing the behavior
of one-loop QCD amplitudes as external momenta become either soft or
collinear.  Our results are valid to all orders in the dimensional
regularization parameter, $\eps$.  Terms through $\Ord(\eps^2)$ can
contribute in infrared divergent phase space integrals associated with
next-to-next-to-leading order jet cross-sections.
\end{abstract}

\end{center}
 \vfil

\end{titlepage}

\section{Introduction}

In recent years, significant progress has been made in computing the
next-to-leading order (NLO) corrections to multi-jet rates within
perturbative QCD.  This progress has come in the form of both matrix
element calculations (see {\it e.g.\/} refs.~\cite{mpReview,bdkReview}) and
algorithms for numerical jet calculations (see
{\it e.g.\/} refs.~\cite{gg,ggk}). An important next step would be the
computation of next-to-next-to-leading-order (NNLO) corrections to
multi-jet rates.  As an example, a calculation of NNLO contributions
to $e^+ e^- \to 3$ jets is needed to further reduce the
theoretical uncertainty in the determination of the QCD coupling,
$\alpha_s$, from event shape variables~\cite{schm}. In fact, many of
the experimental searches for new physics require a precise
understanding of the QCD background, and a more accurate determination
of $\alpha_s$.  At hadron colliders, NNLO calculations would reduce
the dependence of the multi-jet rates on the factorization and
renormalization scales, and would allow a more detailed study of
jet structure.

To compute $n$ particle production at NNLO, three sets of amplitudes
are required: {\it a}) $n$ particle production amplitudes at tree
level, one loop and two loops; {\it b}) $n+1$ particle production
amplitudes at tree level and one loop; {\it c}) $n+2$ particle
production amplitudes at tree level.  For example, the computation of
NNLO $e^+e^-\to3$ jet production requires $e^+e^-\to5$
parton amplitudes at tree level~\cite{ZFiveJetsBorn},
$e^+e^-\to4$ parton amplitudes at tree
level~\cite{ee4tree} and at one loop~\cite{bdkZ4,cgmZ4}, and
$e^+e^-\to3$ parton amplitudes at tree level~\cite{ee3tree}, one
loop~\cite{ert} and two loops.  In terms of the amplitudes required,
the crucial missing piece is the two-loop calculation.  In fact, no
two-loop computations exist for configurations involving more than a
single kinematic variable, except in cases of maximal
supersymmetry~\cite{bry}.

In general, loop-level amplitudes have a complicated analytic
structure, with the complexity increasing rapidly with the
multiplicity of kinematic variables.  The analytic structure
simplifies in the infrared limit (where parton momenta are soft or
collinear) when the amplitudes become singular.  Infrared
singularities exhibit universal, {\it i.e.\/} process independent,
behavior, manifesting themselves as poles in the dimensional regulator
$\eps=(4-D)/2$ after integration over virtual or unresolved momenta.
The Kinoshita-Lee-Nauenberg theorem~\cite{kln} guarantees that the
infrared singularities must cancel for sufficiently inclusive physical
quantities when the real and virtual contributions are combined.  For
processes without strongly interacting initial state particles, like
$e^+ e^- \to X$, the cancellation is complete.  For processes with
strongly-interacting initial states, like jet production in hadron
collisions, initial-state collinear divergences survive the
cancellation and are factorized into the parton distribution
functions, reducing the dependence of the hadron cross section on the
factorization scale, $\mu_F$.

The structure of infrared singularities at NLO is well understood.  In
the squared tree-level amplitudes, any one of the $n+1$ produced
particles can be unresolved in the final state.  The ensuing infrared
singularities are accounted for by tree-level soft~\cite{bcm,bg} and
collinear splitting functions~\cite{ap}.  These have also
been combined into a single function~\cite{dipole}. In addition, the universal
structure of the coefficients of the $1/\eps^2$ and $1/\eps$ poles for
the one-loop virtual contributions to $n$-particle productions is
known~\cite{gg,ks,kstsingular}.  The universality of the structure of
the singularities has been exploited in building general-purpose
algorithms~\cite{gg,ggk,dipole} for NLO jet production calculations.

The study of the infrared structure at NNLO is still underway.  A
detailed understanding of the infrared singularities that arise from
the combination of virtual loops and unresolved real emission will
be crucial to the development of methods for performing NNLO
calculations.  In the squared tree-level amplitudes, any two of the
$n+2$ produced particles can be unresolved.  The resulting soft,
collinear, and mixed soft/collinear singularities are described by
universal tree-level double-soft~\cite{bg}, double-splitting and
soft-splitting~\cite{cg} functions, respectively.  In addition, the
universal structure of the coefficients of the $1/\eps^4$, $1/\eps^3$
and $1/\eps^2$ poles has been determined~\cite{cat} for the two-loop
virtual contributions to $n$-particle production.

In the interference between a one-loop amplitude for $n+1$ particle
production and its tree-level counterpart, unresolved real emission
generates additional $1/\epsilon^2$ divergences, which when combined
with one-loop virtual singularities brings the total order of
divergence to $1/\epsilon^4$.  In order to evaluate the interference
terms to $\Ord(\epsilon^0)$, the $(n+1)$-parton one-loop amplitude
must be evaluated to $\Ord(\epsilon^2)$.  In light of the already
complicated analytic structure of one-loop amplitudes to
$\Ord(\epsilon^0)$~\cite{bdkZ4,cgmZ4,bdk5g,kst1g4q,bdk3g2q,dksWW}, such a
calculation would be a truly formidable task.  A more reasonable
approach is to evaluate the amplitudes to higher order in $\eps$ only
in the infrared regions of phase-space where the amplitudes factorize
into sums of products of $n$-parton final-state amplitudes multiplied
by soft or collinear splitting amplitudes.  It is only these soft and
collinear splitting amplitudes and the $n$-parton final-state one-loop
amplitudes that must be evaluated to higher order in $\eps$. 

In this paper we describe in detail the methods used to obtain the
soft and collinear splitting amplitudes to all orders in $\epsilon$
and present new results for the soft and splitting amplitudes with
quarks, which have previously been given only through
$\Ord(\epsilon^0)$~\cite{bdk3g2q,bddkSusy4}.  We present our results
both in terms of the helicity representation and in terms of formal
polarizations and spinors. The latter form is convenient if one is
working entirely in conventional dimensional
regularization~\cite{CollinsBook}.  In a previous letter~\cite{us}, we
presented all orders in $\epsilon$ results for the one-loop pure gluon
soft and collinear splitting amplitudes and used the one-loop soft
amplitudes to $\Ord(\epsilon)$ to re-derive~\cite{dds2} next-to-leading
logarithmic corrections to the Lipatov vertex~\cite{bfkl}.

In order to present the factorization properties in their simplest
forms, we decompose one-loop QCD amplitudes first into partial
amplitudes~\cite{Color,bkcolor}, which follow from the color structure, and
then into primitive amplitudes~\cite{bdkZ4,bdk3g2q}.  The benefit of this
last step is that color factors are completely disentangled from 
primitive amplitudes, allowing for relatively simple description of the
soft behavior of the amplitudes.

This paper is organized as follows:  In \sec{sec:ColorRev}, we
review the decomposition of one-loop QCD amplitudes by color ordering
and the further decomposition into primitive amplitudes; in
\sec{sec:RevSofCol}, we review the collinear and soft factorization
properties of QCD amplitudes; in \sec{sec:LoopSofCol}, we describe our
method for obtaining the soft and collinear splitting amplitudes to all
orders in $\eps$, which is based upon the discussion of
ref.~\cite{bc}, and present our results; in \sec{sec:Check}, we
check our result by comparing to those obtained from $N=4$
supersymmetric amplitudes (which are known to all orders in
$\eps$~\cite{DimShift}) and those obtained from $gggH$ amplitudes
using the effective $ggH$ coupling~\cite{Dawson} generated by a heavy
fermion loop, again being careful to keep all higher order
in $\eps$ contributions. (For a discussion of this calculation
through $\Ord(\eps^0)$ see ref.~\cite{crs}.)

We note that an interesting technique for obtaining the collinear
splitting amplitudes has been recently proposed~\cite{KosowerSplit},
using the unitarity reconstruction methods of
refs.~\cite{bdkReview,bddkSusy4,bddkSusy1}. This method appears to
allow for a straightforward generalization to higher loops.


\section{Review of color decompositions and primitive amplitudes}
\label{sec:ColorRev}

To discuss the properties of one-loop QCD amplitudes as momenta become
either soft or collinear we will use `primitive amplitudes'.
Primitive amplitudes, defined in refs.~\cite{bdkZ4,bdk3g2q} and
originally motivated by the structure of fermionic string theory
amplitudes, are gauge invariant building blocks from which QCD
amplitudes, including their color factors, can be built.  An important
characteristic of primitive amplitudes is that they have a fixed
ordering of the external legs, leading to a relatively simple
analytic structure.  In traditional representations of QCD amplitudes,
factorization in the collinear limit is fairly straightforward, but in
the soft limit, color factors become entangled with kinematic factors
in a non-trivial way.  At tree level, a conventional color
decomposition~\cite{Color} is sufficient to disentangle the soft
factorization properties.  This is no longer true at one-loop where
the presence of sub-leading color structures results in an even deeper
entanglement of color and kinematics.  Our purpose in using primitive
amplitudes is to provide a clean factorization of one-loop amplitudes
in the soft and collinear regions and to separate color issues from
the kinematic issues.  In \sec{SoftExampleSubSubSection} we will make
use of the formulas reviewed in this section to give an example of how the
simple soft factorization of primitive amplitudes involving quarks
leads to a nontrivial tangle in terms of the more conventional color
ordered partial amplitudes.  In ref.~\cite{us}, where only the pure
glue case was dealt with, there was no need to introduce primitive
amplitudes since the standard leading color partial amplitudes play
the same role.

For the cases of $n$-gluon amplitudes and two quark, $n-2$ gluon
amplitudes general formul\ae\ expressing color ordered partial
amplitudes in terms of primitive amplitudes have been
presented~\cite{bdk3g2q,bddkSusy4}.  For the purposes of this paper we
review this one-loop decomposition for the limited cases of four or
five colored partons, corresponding to NNLO $\bar p p \to 2$ jets,
which will likely be among the first cases to which the results in
this paper can be applied.  We also very briefly outline the construction
of the primitive decomposition of four quark, $n-4$ gluon amplitudes.  The
primitive decomposition for the one-loop contributions to $e^+ e^- \to
3$ jets and $e^+ e^- \to4$ jets may be found in ref.~\cite{bdkZ4}.

The usual color decomposition follows from replacing the color
structure constants with commutators of fundamental representation
matrices,%
\footnote{Note that we use a non-standard normalization for
fundamental representation matrices, $\Tr[T^aT^b] = \delta^{ab}$.}
\begin{equation}
f^{abc} = -{i \over \sqrt{2}} (\Tr[T^a T^b T^c] - \Tr[T^b T^a T^c]) \,,
\end{equation}
and then using $SU(N_c)$ Fierz identities,
\begin{equation}
T^a_{i \bar \jmath} \, T^a_{m \bar n} =  \delta_{i \bar n} \,\delta_{m \bar
\jmath}  - {1\over N_c}\, \delta_{i \bar \jmath} \, \delta_{m \bar n}\,,
\end{equation}
to combine traces.  In the case of amplitudes with purely adjoint
particles one may use instead the $U(N_c)$ Fierz identities which do
not contain the second self contraction term. 

At tree level the pure gluon amplitude may be expressed 
as~\cite{mpReview,Color}
\begin{equation}
\A{n}^{\tree}(1,2,\ldots, n) = g^{n-2} \!
\sum_{\sigma \in S_n/Z_n} \!
\Tr(T^{a_{\si(1)}}T^{a_{\si(2)}} \cdots T^{a_{\si(n)}}) \,
A_n^{\tree} (\si(1), \si(2), \ldots, \si(n)) \,,
\label{TreeGluonDecomp}
\end{equation}
where $S_n/Z_n$ is the set of all non-cyclic permutations.
For the case of two quarks in the fundamental representation 
and $(n-2)$ gluons in the adjoint representation the color decomposition is
\begin{equation}
 {\cal A}_n^\tree(1_{\bar{q}},2_q,3,\ldots,n)
  =   g^{n-2} \sum_{\sigma\in S_{n-2}}
   (T^{a_{\sigma(3)}}\ldots T^{a_{\sigma(n)}})_{i_2}^{~\ib_1}\
    A_n^\tree(1_{\bar{q}},2_q;\sigma(3),\ldots,\sigma(n))\, ,
\label{TwoQuarkGluonDecomp}
\end{equation}
where $S_{n-2}$ is the permutation group on $n-2$ elements.
Similar expressions exist for the case of larger numbers of fermions.
In each case, the terms which multiply different color
structures are called `partial amplitudes'. 

We may also define a set of tree-level primitive amplitudes for the
pure gluon case to correspond to the partial amplitudes.  For the case
of two fermions the primitive amplitudes are the set of partial
amplitudes one would obtain if all particles, including the fermions,
were in the adjoint representation.

This decomposition yields color structures consisting of products of
fundamental representation matrices.  In the simplest color
structures the product involves a single chain of representation
matrices.  Tree-level amplitudes only use the leading color
structures.  There are also more complicated (sub-leading) color
structures, which come in one-loop, in which the product breaks into
more than one chain.  

For example, the color decomposition of the one-loop four-gluon
amplitude is
\begin{equation}
\eqalign{
\A{4}^{\oneloop}(1,2,3,4) = g^4\left[\vphantom{\sum_{\sigma \in P_4}}\right.
&\sum_{\sigma \in S_4/Z_4} 
N_c \Tr(T^{a_{\si(1)}}T^{a_{\si(2)}}T^{a_{\si(3)}}T^{a_\si{(4)}})
A_{4;1}^{[1]} (\si(1), \si(2), \si(3), \si(4)) \cr
 + &\sum_{\si \in S_4/Z_2^3} \Tr(T^{a_{\si(1)}}T^{a_{\si(2)}})
\Tr(T^{a_{\si(3)}}T^{a_{\si(4)}})
A_{4;3} (\si(1), \si(2); \si(3), \si(4)) \;.\cr
& + n_{\! f} \sum_{\sigma \in S_4/Z_4}
\Tr(T^{a_{\si(1)}}T^{a_{\si(2)}}T^{a_{\si(3)}}T^{a_\si{(4)}})
A_{4;1}^{[1/2]}(\si(1), \si(2), \si(3), \si(4)) \cr
& + n_{s} \left.\sum_{\sigma \in S_4/Z_4}
\Tr(T^{a_{\si(1)}}T^{a_{\si(2)}}T^{a_{\si(3)}}T^{a_\si{(4)}})
A_{4;1}^{[0]}(\si(1), \si(2), \si(3), \si(4))\right] \,.\cr}
\label{Part4Glue}
\end{equation}
In the first term, the permutation $\sigma$ lies in the set of all
permutations ($S_4$) of four objects with purely cyclic ones ($Z_4$)
removed. In the second term, $\si$ is again in the set of permutations
of four objects but with two factors of $Z_2$ removed, corresponding
to exchanging the indices within each trace, as well as another $Z_2$
removed corresponding to interchanging the two traces.  The leading
partial amplitude is indicated by the subscript `4;1', and the
sub-leading partial amplitude is indicated by the subscript `4;3'.  In
this process, only a single sub-leading color structure appears.
Processes with more external legs can have more sub-leading color
structures.  Note that in \eqn{Part4Glue}, we have
abbreviated the dependence of the $A_{n;j}$ on momentum $k_l$ and
helicity $\lambda_l$ by writing the label $l$ alone.

The superscripts $[1]$, $[1/2]$ and $[0]$ label the spin of the
particles circulating in the loops.  The third and fourth terms are
part of the leading partial amplitude and correspond to the
contributions of $n_f$ flavors of fundamental representation quarks
and $n_s$ flavors of complex scalars in the $(N_c + \overline{N}_c)$
representation.  In QCD, below the top threshold $n_f = 5$ and $n_s=0$.
Although fundamental scalars do not appear in QCD, it is convenient to
keep explicit dependence on the number of scalars; when the number of
bosonic states are equal to the number of fermion states certain
supersymmetric Ward identities~\cite{SWI} must be respected which can
be used as a check on results.

In our convention, each matter representation has $4N_c$ color and
helicity degrees of freedom.  This does not correspond to the standard
assignment of $2N_c$ degrees of freedom for scalars, but has the
advantage of making supersymmetry identities more apparent since the
number of states matches that of a Dirac fermion.

Similarly the five-point amplitude is,
\begin{equation}
\eqalign{
{\cal A}_{5}^\oneloop = & g^5 \left[
\sum_{\sigma \in S_5/Z_5}
N_c \Tr(\si(1)\ldots\si(5))
    A_{5;1}^{[1]} (\si(1),\ldots,\si(5))\right. \cr
& + \hskip-4mm
  \sum_{\sigma \in S_5/(Z_2\times Z_3)} \hskip-6mm
  \Tr(\si(1)\si(2)) \Tr(\si(3)\si(4)\si(5))
A_{5;3} (\si(1),\si(2);\si(3),\si(4),\si(5)) \cr
& + 
n_{\! f} \sum_{\sigma \in S_5/Z_5}
   \Tr(\si(1)\ldots\si(5))
    A_{5;1}^{[1/2]} (\si(1),\ldots,\si(5)) \cr
& + \left.
n_s \sum_{\sigma \in S_5/Z_5}
\Tr(\si(1)\ldots\si(5))
    A_{5;1}^{[0]}(\si(1),\ldots,\si(5)) \right] \,, \cr}
\label{LoopColor}
\end{equation}
where $S_5/Z_5$ is the set of non-cyclic permutations of five objects
and $S_5/(Z_2\times Z_3)$ is the set of permutations of five objects
that do not leave the product of two and three element traces
invariant. 

In the pure external gluon case the leading color amplitudes
(broken down by the spin of the loop particle) play the role of
primitive amplitudes in the sense that all sub-leading partial
amplitudes can be obtained by the appropriate permutation sum
given by 
\begin{equation}
A_{n;c>1}(1,2,\ldots,c-1;c,c+1,\ldots,n)\ =\
 (-1)^{c-1} \sum_{\sigma\in COP\{\alpha\}\{\beta\}}
 A^{[1]}_{n;1}(\sigma) \,,
\label{sublanswer}
\end{equation}
where $\alpha_i \in \{\alpha\} \equiv \{c-1,c-2,\ldots,2,1\}$,
$\beta_i \in \{\beta\} \equiv \{c,c+1,\ldots,n-1,n\}$,
and $COP\{\alpha\}\{\beta\}$ is the set of all
permutations of $\{1,2,\ldots,n\}$ with $n$ held fixed
that preserve the cyclic
ordering of the $\alpha_i$ within $\{\alpha\}$ and of the $\beta_i$
within $\{\beta\}$, while allowing for all possible relative orderings
of the $\alpha_i$ with respect to the $\beta_i$.
For example if $\{\alpha\} = \{2,1\}$ and
$\{\beta\} = \{3,4,5\}$, then $COP\{\alpha\}\{\beta\}$
contains the twelve elements
\begin{equation}
\eqalign{
 &(2,1,3,4,5),\quad (2,3,1,4,5),\quad (2,3,4,1,5),\quad
  (3,2,1,4,5),\quad (3,2,4,1,5),\quad (3,4,2,1,5), \cr
 &(1,2,3,4,5),\quad (1,3,2,4,5),\quad (1,3,4,2,5),\quad
  (3,1,2,4,5),\quad (3,1,4,2,5),\quad (3,4,1,2,5) \,. \cr}
\end{equation}

The color decomposition of the $\bar{q}qgg$ one-loop amplitude is
\cite{bdk3g2q,kstfour}
\begin{equation}
\eqalign{
 \A{4}(1_{\bar{q}},2_q,3,4) = g^4 \left[
      N_c\,\sum_{\sigma\in S_2}\right.&(T^{a_{\si(3)}}
              T^{a_{\si(4)}})_{i_2}^{~\ib_1}
              \ A_{4;1}(1_{\bar{q}},2_q;\si(3),\si(4)) \cr
+&\left.\vphantom{\sum_{\sigma\in S_2}}
    \Tr(T^{a_3}T^{a_4}) \, \delta_{i_2}^{~\ib_1}
              \ A_{4;3}(1_{\bar{q}},2_q;3,4)\right] \,.\cr}
\label{qqggDecomp}
\end{equation}
Similarly, the color decomposition of the $\bar{q}qggg$ amplitude
is~\cite{kstsingular,bdk3g2q}
\begin{equation}
\eqalign{
 \A{5}(1_{\bar{q}},2_q,3,4,5) = g^5 \left[
      N_c\, \sum_{\sigma\in S_3}\right.
 (T^{a_{\sigma(3)}}T^{a_{\sigma(4)}}T^{a_{\sigma(5)}})_{i_2}^{~\ib_1}
    \ &A_{5;1}(1_{\bar{q}},2_q;\sigma(3),\sigma(4),\sigma(5)) \cr
  + \sum_{\sigma\in Z_3} \Tr\L T^{a_{\sigma(3)}} T^{a_{\sigma(4)}}\R\;
              \L T^{a_{\sigma(5)}}\R_{i_2}^{~\ib_1}
              \  &A_{5;3}(1_{\bar{q}},2_q;\sigma(3),\sigma(4);\sigma(5))\cr
   + \sum_{\sigma\in S_2} \Tr\L T^{a_{\sigma(3)}}T^{a_4}T^{a_{\sigma(5)}}\R \, \delta_{i_2}^{~\ib_1}
              \  &A_{5;4}(1_{\bar{q}},2_q;\sigma(3),4,\sigma(5))
              \left.\vphantom{\sum_{\sigma\in Z_3}} \right]\,.\cr}
\label{qqgggDecomp}
\end{equation}
In the partial amplitude $A_{5;3}$ an additional semicolon separates
the gluon sandwiched between the quark indices
(the last gluon in $A_{5;3}$) from the other two gluons.
 
For the case of external fermions the leading partial amplitudes
cannot play the role of primitive amplitudes.  The primitive amplitudes 
with nearest neighboring quarks are
given by subdividing $A_{n;1}$ into smaller pieces~\cite{bdk3g2q},
\begin{equation}
\eqalign{
  A_{n;1}(1_{\bar{q}},2_q;3,\ldots,n) &\equiv
   A_n^{L,\,[1]}(1_{\bar{q}},2_q,3,\ldots,n)
   - {1\over N_c^2} A_n^{R,\,[1]}(1_{\bar{q}},2_q,3,\ldots,n) \cr
 &\quad + {\nf\over N_c} A_n^{L,\,[1/2]}(1_{\bar{q}},2_q,3,\ldots,n)
  + {\ns\over N_c} A_n^{L,\,[0]}(1_{\bar{q}},2_q,3,\ldots,n) \,. \cr}
\label{Anoneformula}
\end{equation}
All color factors have been extracted into the coefficients of the
primitive amplitudes.  The primitive amplitudes themselves are independent
of the numbers of colors.  The labels $L$ and $R$ on the 
primitive amplitudes refer to whether the fermion line which 
enters a diagram turns either `left' or `right'.  For the purposes in
this paper we may take \eqn{Anoneformula} to be the defining 
equation for the two-quark $n-2$ gluon amplitudes, when all gluons 
are between quark and the anti-quark in the cyclic ordering.
Further details, and the definition of the primitive amplitudes
when gluons are also between the anti-quark and quark, 
may be found in ref.~\cite{bdk3g2q}.  The $L$ and $R$ primitive
amplitudes are related by an inversion of the ordering of legs,
\begin{equation}
A_n^{L,\,[J]}(1_{\bar{q}},2,\ldots,k-1,k_q,k+1,\ldots,n) = (-1)^n
A_n^{R,\,[J]}(1_{\bar{q}},n,\ldots,k+1,k_q,k-1,\ldots,2)\,.
\label{LandR}
\end{equation}
For the $L,[1]$ primitive amplitudes all gluon legs between the anti-quark 
and the quark in the
cyclic ordering are attached to the fermionic part of the loop,
while all gluon legs between the quark and the anti-quark are 
attached to the bosonic part of the loop.  (See ref.~\cite{bdk3g2q} for 
further details.) 

The sub-leading partial amplitudes of two quark processes are given by
permutation sums similar to \eqn{sublanswer} for all-gluon processes,
\begin{equation}
\eqalign{
A_{n;c>1}(1_\qb,2_q;3,\ldots,c+&1;c+2,\ldots,n) = \cr
 &(-1)^{c-1}\sum_{\sigma\in COP\{\alpha\}\{\beta\}}
 \left[A^{L,\,[1]}_{n;1}(\sigma)
  - {n_f\over N_c}A^{R,\,[1/2]}_{n;1}(\sigma)
  - {n_s\over N_c}A^{R,\,[0]}_{n;1}(\sigma)\right] \,,\cr
}
\label{subl2quark}
\end{equation}
where $\alpha_i \in \{\alpha\} \equiv \{c+1,c,\ldots,4,3\}$,
$\beta_i \in \{\beta\} \equiv \{1_\qb,2_q,c+2,c+3,\ldots,n-1,n\}$,

For example, for the four-point $\qb qgg$ amplitude,
\begin{equation}
\eqalign{
A_{4;3}(1_\qb, 2_q ; 3, 4)\ =\
  &A_4^{L,\,[1]}(1_\qb, 2_q, 3, 4)
 + A_4^{L,\,[1]}(1_\qb, 2_q, 4, 3)
 + A_4^{L,\,[1]}(1_\qb, 3, 2_q, 4)  \cr
 +&A_4^{L,\,[1]}(1_\qb, 3, 4, 2_q)
 + A_4^{L,\,[1]}(1_\qb, 4, 3, 2_q)
 + A_4^{L,\,[1]}(1_\qb, 4, 2_q, 3) \,, \cr}
\label{Afourthree}
\end{equation}
Symmetry relations among $\nf$ and $\ns$ terms cause them to cancel
out.

The five-point relations are, of course, a bit more complicated,
\begin{equation}
\eqalign{
A_{5;3}(1_\qb, 2_q ; 4,5; 3) =&\sum_{\si\in S_3}  
       A_5^{L,\,[1]}(1_\qb, 2_q, \si(3), \si(4), \si(5))
     + A_5^{L,\,[1]}(1_\qb, 4, 2_q, 5, 3) \cr
     +&A_5^{L,\,[1]}(1_\qb, 4, 2_q, 3, 5)
     + A_5^{L,\,[1]}(1_\qb, 5, 2_q, 3, 4)
     + A_5^{L,\,[1]}(1_\qb, 5, 2_q, 4, 3) \cr
     +&A_5^{L,\,[1]}(1_\qb, 4, 5, 2_q, 3)
     + A_5^{L,\,[1]}(1_\qb, 5, 4, 2_q, 3)\, , \cr}
\label{AfivethreeA}
\end{equation}
\begin{equation}
\eqalign{
A_{5;4}(1_\qb, 2_q ; 3, 4,5) = \sum_{\sigma \in Z_3}
\Biggl[
- &     A_5^{L,\,[1]}(1_\qb, 2_q , \sigma(5), \sigma(4), \sigma(3))
      - A_5^{L,\,[1]}(1_\qb,  \sigma(5), 2_q, \sigma(4), \sigma(3)) \cr
&     - A_5^L(1_\qb, \sigma(5), \sigma(4), 2_q, \sigma(3))
      - A_5^L(1_\qb,  \sigma(5), \sigma(4), \sigma(3), 2_q)  \cr
  - {\nf\over N_c} \Bigl(
       &A_5^{L,\,[1/2]}(1_\qb, 2_q , \sigma(3), \sigma(4), \sigma(5))
      + A_5^{L,\,[1/2]}(1_\qb, \sigma(3), 2_q, \sigma(4), \sigma(5)) \Bigr) \cr
  - {\ns\over N_c}          \Bigl(
       &A_5^{L,\,[0]}(1_\qb, 2_q , \sigma(3), \sigma(4), \sigma(5))
      + A_5^{L,\,[0]}(1_\qb, \sigma(3), 2_q, \sigma(4), \sigma(5)) \Bigr)
\Biggr] \,. \cr}
\label{AfivefourA}
\end{equation}

The primitive decomposition of four quark amplitudes have not been
explicitly given in the literature.  Nevertheless, such a
decomposition can be performed by following the ideas in ref.~\cite{bdk3g2q}.
The first step in performing a decomposition into primitive amplitudes
is to perform the usual decomposition into coefficients of independent
color factors.  Then as in the previous cases one defines a set of
primitive amplitudes in terms of a set of `parent' diagrams.  As for
the other cases, the essential distinguishing feature of the primitive
amplitudes is that the colored external legs of all contributing
diagrams have a fixed cyclic ordering.  It is this general feature of
primitive amplitudes which we use in subsequent sections to present a
systematic description of the soft properties of the one-loop
amplitudes.


\section{Review of collinear and soft properties of QCD amplitudes}
\label{sec:RevSofCol}

The properties of QCD tree amplitudes in the limits where momenta
become collinear or soft has been extensively discussed in the
literature.  (See for example ref.~\cite{mpReview}.)  Similarly, the
properties of one-loop amplitudes in collinear limits have also been
extensively discussed~\cite{bdkReview,bdk3g2q,bddkSusy4,bc} through
$\Ord(\eps^0)$. Although, the properties of one-loop QCD amplitudes as
external momenta become soft has been less extensively discussed,
it is straightforward to extract explicit
expressions for the behavior using the known
four-point~\cite{kstfour,bk} and five-point 
amplitudes~\cite{bdk5g,kst1g4q,bdk3g2q}.  
In this paper we wish to
extend these results to higher orders in the dimensional
regularization parameter, so that they can be inserted into NNLO
computations.

In addition to the application of using the soft and collinear splitting
amplitudes to evaluate the infrared divergent regions of phase space
integrals, the understanding of the collinear and soft properties of
one-loop amplitudes has led to a method for constructing amplitudes
from their analytic properties.  This has been used to simplify the
computation of the $Z \to 4$ parton helicity amplitudes and to obtain
infinite sequences of one-loop maximally helicity violating
amplitudes~\cite{bdkReview,bddkSusy4,bcdkAllPlus}.

We now review the known soft and collinear properties of the one-loop
amplitudes in preparation for our extension of the results to all
orders of the dimensional regularization parameter.

\subsection{The collinear behavior}

\begin{figure}[ht]
\centerline{\epsfxsize 3.5 truein \epsfbox{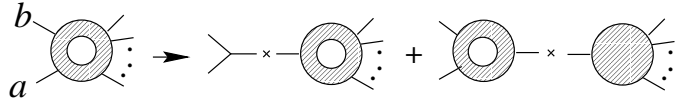}}
\vskip -.2 cm
\caption[]{
\label{CollFactFigure}
\small The generic behavior of one-loop amplitudes in the limit as two
external momenta become collinear.  The shaded disc represents the sum
over tree diagrams and the annulus the sum over one-loop diagrams.}
\end{figure}

As the momenta of two massless particles become collinear, one-loop
amplitudes factorize in the way depicted in \fig{CollFactFigure}.  As
discussed in ref.~\cite{bc}, for infrared divergent massless theories
the situation is more subtle than in infrared finite cases since the
various contributions to individual integral functions diagrams may
not have a smooth behavior as the intermediate momentum becomes
massless.  In particular, in terms of the diagrams of covariant gauges, the
picture of \fig{CollFactFigure} is slightly misleading since 
there can be contributions from diagrams with no
explicit propagator on which to factorize the amplitude; the
kinematic poles arise instead from the loop integral.  (This subtlety
is related to the interchange of the limit $\eps \to 0$ with that of
a kinematic variable vanishing $s_{ab} \to 0$.) Nevertheless, the
complete amplitudes do obey the factorization described by
\fig{CollFactFigure}.

%

More explicitly, as two adjacent momenta $k_1$ and $k_2$ become collinear, 
we may factorize an $n$-point tree amplitude  on the $s_{12}$ kinematic 
pole in terms of a three vertex and an $(n-1)$-point amplitude, 
\begin{equation}
\eqalign{
  A_n^{\tree} (1,2, \ldots, n) & \iscol{1}{2}
   {\cal D}^{\mu,\,\tree}_{K\to1,2}
    \left(i\sum_{\lambda}{\pol^{-\lambda}_\mu(K)\pol^{\lambda}_\nu(K)
    \over s_{12}}\right)
    {\partial\over\partial\pol^\lambda_\nu(K)}
    A^{\tree}_{n-1}(K^\lambda, 3, \dots)\,,\cr }
\label{FormalTreeCollinear}
\end{equation}
where $\lambda$ specifies the polarization and $k_1=zK$ and $k_2=(1-z)K$
with $K = k_1 + k_2$.  This basic structure is independent of
the particle type, although we have written \eqn{FormalTreeCollinear}
for the case of an intermediate gluon. The values of the
${\cal D^\tree}$ functions do however depend on the particle types and
are simply given by the three-vertices
\begin{equation}
\eqalign{
{\cal D}^{\mu,\,\tree}_{g \to g_1g_2} &= i\sqrt{2} (k_1^\mu \pol_1\cdot \pol_2 
  + k_2\cdot\pol_1 \pol_2^\mu - k_1\cdot\pol_2 \pol_1^\mu) \,,\cr 
{\cal D}^{\mu,\,\tree}_{g \to \qb_1 q_2} & = 
{i \over \sqrt{2} } \overline{u}_2 \gamma^\mu v_1 \,,\cr
{\cal D}^{j,\,\tree}_{q \to q_1g_2} & = 
{i\over\sqrt{2}}\overline{u}_{j1}\feynsl{\pol}_2\,.
               \cr}
\label{TreeThreeVertices}
\end{equation}
In the first two cases $\mu$ represents
the Lorentz index of the intermediate gluon which in the last case
is replaced by a spinor index.  Although it makes no difference
in the results, we use the non-linear Gervais-Neveu~\cite{gn} gauge since the
vertices are particularly simple.

%
\begin{figure}[ht]
\centerline{\epsfxsize 3.5 truein \epsfbox{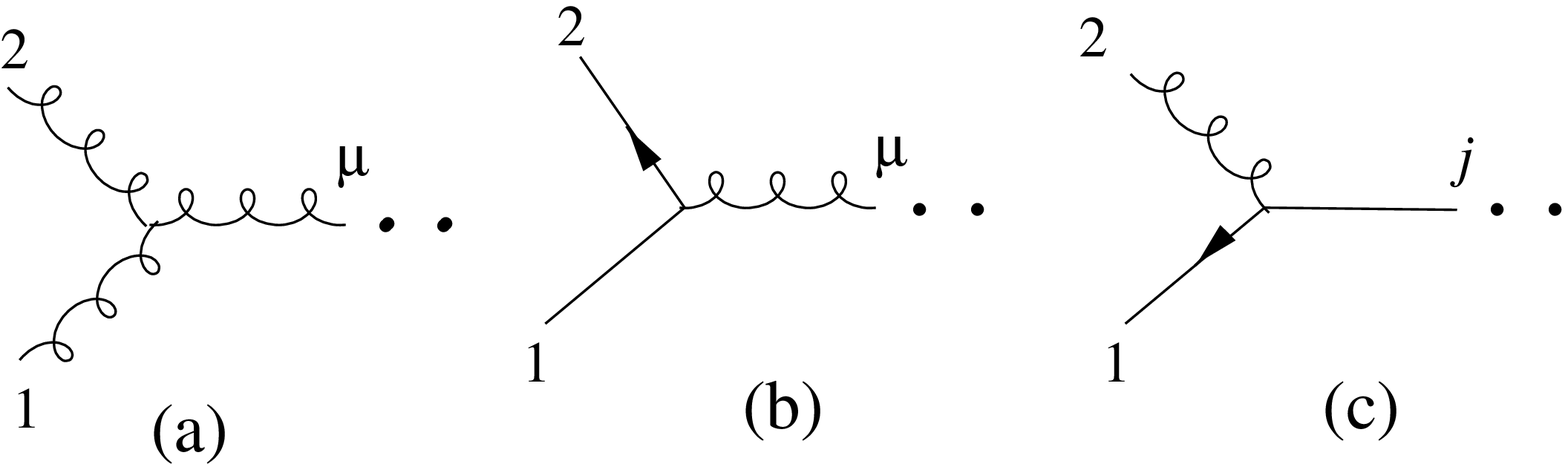}}
\vskip -.2 cm
\caption[]{
\label{ThreeVertexFigure}
\small The three vertices describing (a) ${\cal D}^{\mu,\,\tree}_{g
  \to g_1g_2}$, (b) ${\cal D}^{\mu,\,\tree}_{g \to \qb_1 q_2}$ and (c)
  ${\cal D}^{j,\,\tree}_{q \to q_1g_2}$.} 
\end{figure}

After inserting an explicit representation of the helicity
states~\cite{SpinorHelicity}, the collinear limits of tree amplitudes
may be re-expressed as
(see {\it e.g.\/} ref.~\cite{mpReview})
\begin{equation}
A_{n}^{\tree}(1^{\lambda_1},2^{\lambda_2},\dots, n)
\mathop{\longrightarrow}^{1 \parallel 2}\ 
\sum_{\lambda=\pm} \Split^{\tree}_{-\lambda}
(1^{\lambda_1},2^{\lambda_2})\,
      A_{n-1}^{\tree}(K^\lambda, 3, \ldots, n) \,.
\label{TreeCollinear}
\end{equation}  
The splitting amplitudes in \eqn{TreeCollinear} have square-root
singularities in the collinear limit.  For convenience we have
collected the tree-level helicity splitting amplitudes in
\app{app:TreeSplitAppendix}.  For most computational purposes, the
helicity form in \eqn{TreeCollinear} is the more convenient one, but
if one is working entirely in the context of conventional dimensional
regularization where the polarization vectors become
$(4-2\eps)$-dimensional, the representation in terms of formal
polarization vectors and spinors is also useful.

The behavior of one-loop primitive amplitudes as the momenta of
two adjacent legs becomes collinear is similar and
is given by
\begin{equation}
\eqalign{
A_{n}^{\oneloop}(1,2,\dots, n)
&\mathop{\longrightarrow}^{1 \parallel 2}\ 
    {\cal D}^{\mu,\,\tree}_{K\to1,2}
    \left(i\sum_{\lambda}{\pol^{-\lambda}_\mu(K)\pol^{\lambda}_\nu(K)
    \over s_{12}}\right)
    {\partial\over\partial\pol^\lambda_\nu(K)}
    A^{\oneloop}_{n-1}(K^\lambda, 3, \dots)\cr
& \hskip 1cm
  +  {\cal D}^{\mu,\,\oneloop}_{K\to1,2}
    \left(i\sum_{\lambda}{\pol^{-\lambda}_\mu(K)\pol^{\lambda}_\nu(K)
    \over s_{12}}\right)
    {\partial\over\partial\pol^\lambda_\nu(K)}
    A^{\tree}_{n-1}(K^\lambda, 3, \dots)\,.\cr
}
\label{FormalOneLoopCollinear}
\end{equation}
In the spinor helicity language, this becomes~\cite{bddkSusy4,bdk3g2q}
\begin{equation}
\eqalign{
A_{n}^{\oneloop}(1^{\lambda_1},2^{\lambda_2},\dots, n)
&\mathop{\longrightarrow}^{1 \parallel 2}\ 
\sum_{\lambda=\pm}  \biggl\{
\Split^{\tree}_{-\lambda}
(1^{\lambda_1},2^{\lambda_2})\,
      A_{n-1}^{\oneloop}(K^\lambda, 3,\ldots) \cr
& \hskip 2 cm 
+\Split^{\oneloop}_{-\lambda}
(1^{\lambda_1},2^{\lambda_2})\,
      A_{n-1}^{\tree}(K^\lambda,3,\ldots) \biggr\} \,, \cr}
\label{OneLoopCollinear}
\end{equation}
where $\lambda$ represents the helicity, $A_{n-1}^{\oneloop}$ and
$A_n^{\tree}$ are one-loop and tree sub-amplitudes with a fixed
ordering of legs with legs $1$ and $2$ consecutive in the ordering.

In \sec{sec:LoopSofCol} we give expressions for the one-loop splitting
functions ${\cal D}^{\mu,\,\oneloop}$ and $\Split^{\oneloop}$ to all
orders in the dimensional regularization parameter.  

\subsection{Soft behavior}

In the limit that a gluon momentum becomes soft
we may again factorize tree amplitudes on kinematic
poles (see {\it e.g.\/} ref.~\cite{mpReview}). 
In taking $k_1 \to 0$ the two kinematic 
poles in the color ordered tree amplitude which may diverge are
$1/s_{12}$ and $1/s_{n1}$ since $n$, $1$ and $2$ are consecutive legs in 
the cyclic color ordering.  This yields the factorization relation
\begin{equation}
A_{n}^\tree (1^\pm, 2, \ldots, n)
\mathop{\longrightarrow}^{k_1 \to 0}
 \Soft^{\tree}(n,1^\pm, 2)\, A_{n-1}^\tree(2, 3, \ldots, n) \,. 
\label{TreeSoft}
\end{equation}
In terms of polarization vectors, the tree-level soft function,
$\Soft^{\tree}$, is the familiar eikonal factor,
\begin{equation}
\Soft^{\tree}(n,1^\pm, 2) = -{1\over \sqrt{2}}
      \left[{2\pol^\pm_1\cdot k_n\over s_{n1}} -
       {2\pol^\pm_1\cdot k_2 \over s_{12}}\right] \,.
\label{Eikonal}
\end{equation}
In spinor helicity notation, the soft amplitudes are, 
\begin{equation}
\Soft^\tree(n,1^+,2) = { \spa{n}.2 \over \spa{n}.1 \spa1.2}\,,
\hskip 2 cm 
\Soft^\tree(n,1^-,2) = {-\spb{n}.2 \over \spb{n}.1 \spb1.2}\,.
\label{HelSoft}
\end{equation}
The soft amplitudes are independent of the helicities and particle types
of the neighboring legs $n$ and $2$.

The behavior of one-loop gluon amplitudes in the soft limit is similar to
the collinear behavior.  As the momentum $k_1$ becomes soft the
primitive one-loop amplitudes become
\begin{equation}
A_{n}^\oneloop (1,2, \ldots, n)
 {{\buildrel k_1 \to 0 \over\longrightarrow}}
\Soft^{\tree}(n,1^\pm, 2)\, A_{n-1}^\oneloop(2,3,\ldots, n) 
+ \Soft^\oneloop (n,1^\pm,2)\, A_{n-1}^\tree(2,3, \ldots, n)\,. 
\label{OneLoopSoft}
\end{equation}
The one-loop soft function may be extracted from four- \cite{kstfour,bk,es} and
five-parton \cite{bdk5g,kst1g4q,bdk3g2q} one-loop amplitudes that are
known through $\Ord(\eps^0)$, with the result,
\begin{equation}
\Soft^{\oneloop}(n,1^\pm,2) = - \Soft^\tree(n,1^\pm,2)\,
c_{\Gamma}\, \Bigl[ {1\over\epsilon^2}\,
\left({\mu^2(-s_{n2})\over (-s_{n1})(-s_{12})}\right)^{\epsilon}\,
+ {\pi^2 \over 6} \Bigr] \,.
\label{LeadingSoft}
\end{equation}
As for the collinear case, in \sec{sec:LoopSofCol}, we will generalize
this to all orders in the dimensional regularization parameter.  We
will also generalize the discussion to include the case of one-loop
primitive amplitudes with external fermions and present a
representation in terms of formal polarization vectors. 
As we shall see, although the soft function does not depend on the
neighboring external particles in the primitive ordering of legs, it
does depend on whether the soft gluon is attached to a gluonic or
fermionic part of the loop.


\section{One-loop soft and collinear splitting amplitudes}
\label{sec:LoopSofCol}

In this section, we will apply the analysis of collinear limits and
their relationship to infrared divergences of ref.~\cite{bc} to obtain
explicit expressions for the splitting amplitudes to all orders in the
dimensional regulator, $\eps$.  We shall also present explicit
expressions for the soft amplitudes to all orders in $\eps$.

\begin{figure}[ht]
\centerline{\epsfxsize 3.4 truein \epsfbox{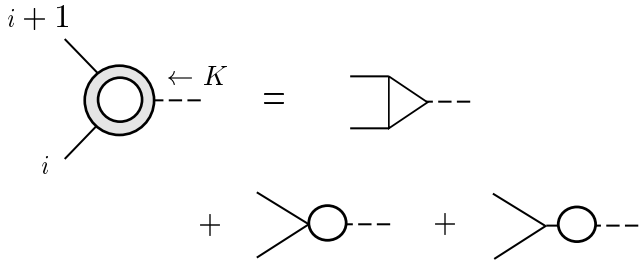}}
\vskip -.2 cm
\caption[]{
\label{MasterSplitLoopFigure}
\small The diagrams in a massless theory (ignoring tadpoles) that need
to be calculated to obtain the factorizing contribution to the loop
splitting function.  The dotted line represents the off-shell leg on
which the collinear factorization is performed.}
\end{figure}

In our organization of the results, there are two contributions to the
one-loop splitting amplitudes; the `factorizing' and the
`non-factorizing' pieces,
$\Split^{\oneloop}=\Split^{\fact}+\Split^{\nonfact}$.  The factorizing
contributions are those which one na\"{\i}vely expects for one-loop
splitting. That is, they are derived from those diagrams, shown in
\fig{MasterSplitLoopFigure}, in which one particle from a tree-level
amplitude splits into two via a loop.  They are called factorizing
contributions because they factorize on a single particle propagator;
one can amputate the splitting term from the diagram by cutting a
single line.

The non-factorizing contributions are quite different in nature.
These arise from non-smooth behavior of infrared divergent integrals
as a kinematic invariant vanishes.  This non-smooth behavior arises
because we take $\eps \rightarrow 0$ before we take the kinematic
invariant to vanish.  These non-factorizing contributions are
proportional to the tree-level ones since the infrared singular parts
of one-loop amplitudes are proportional to the tree-level amplitudes.

Although the division of the splitting amplitudes into factorizing and
non-factorizing pieces contains some arbitrariness and is gauge
dependent, the sum of the contributions is gauge independent.  The
calculation of the diagrams in \fig{MasterSplitLoopFigure} always
gives the complete non-singular contribution to the one-loop
splitting amplitudes, but, depending on the gauge choice, also gives
some portion of the infrared singular contribution.  However, since
the divergence structure of one-loop splitting amplitudes is
known (see \app{app:LoopSplitAppendix} for their values through 
$\Ord(\eps^0)$) and since each $1/\eps$ pole is
uniquely identified with a known `discontinuity function' (see
reference~\cite{bc} and \sec{sec:nonfact} below) we can push all
singularities into the non-factorizing piece, and define the
factorizing piece to be the infrared finite part of the diagrams in
\fig{MasterSplitLoopFigure}, not involving any of the discontinuity
functions.  The non-factorizing contribution may then be determined
by finding the linear combination of discontinuity functions that
matches the known infrared divergence structure of the amplitudes.

We will first compute the diagrams that yield the factorizing parts
of the one-loop splitting amplitudes.   We then use the
known structure of the infrared singularities of the leading partial
amplitudes to derive the non-factorizing contributions to one-loop
splitting.  By combining the factorizing and non-factorizing
parts we obtain the full one-loop splitting amplitudes for the
leading partial amplitudes.  Finally, we will derive the soft factorization
properties of one-loop amplitudes.  Our results will be in terms 
of the bare splitting amplitudes; in \sec{BareToRenormalizedSubSubSection}
we give the appropriate ultraviolet subtraction for obtaining the 
splitting amplitudes for the renormalized amplitudes.

\subsection{Factorizing contributions to the one-loop splitting amplitudes}

The factorizing contributions to one-loop splitting amplitudes are
determined by computing triangle and bubble diagrams like
those shown in \fig{MasterSplitLoopFigure}.  Any poles in $\eps$ that
appear must come from one of two discontinuity functions~\cite{bc}, 
\begin{equation}
\eqalign{
  f_1(s_{12},\eps)& ={1\over\eps(1-2\eps)}
    \left({\mu^2\over-s_{12}}\right)^\eps\,,\cr
  f_2(s_{12},\eps)&= -{1\over\eps^2}
    \left({\mu^2\over-s_{12}}\right)^\eps\,.\cr}
\label{disconts}
\end{equation}
After removing the poles by subtracting off terms proportional to
these discontinuity functions the factorizing contribution to the
splitting function is determined by inserting a complete set of states
on the (off-shell) fused leg, imposing the collinear limit and then
taking the fused leg on-shell.

The results below are computed in the Feynman background field gauge
and apply to an $SU(N_c)$ gauge theory with matter content of $n_f$
flavors of massless Dirac fermions in the fundamental representation
and $n_s$ flavors of massless complex scalars in the $(N_c +
\overline{N}_c)$ representation.  In each case, we describe a parton
splitting into two final state partons labeled $1$ and $2$ which carry
momentum fractions $z$ and $1-z$ respectively.

The difference between dimensional regularization schemes is
parameterized by the quantity $\delta_R$.  In the dimensional
reduction~\cite{dimred} or four-dimensional helicity~\cite{bk}
schemes used to compute one-loop helicity amplitudes, both external
gluons and those inside loops, have two polarization states.  In these
schemes, $\delta_R=0$.  In the 't~Hooft--Veltman scheme, external
gluons have two polarization states, but those inside loops have
$[\eps]$-helicities, {\it i.e.\/} polarization states that point into
the extra $-2\eps$ dimensions.  For this scheme, $\delta_R=1$.  In
conventional dimensional regularization all gluons, external and
internal, have $[\eps]$-helicities.  The $[\eps]$-helicities of the
internal gluons are accounted for by setting $\delta_R=1$; those of
the external gluons must be accounted for by spin sums over
$D$-dimensional polarization vectors.

\subsubsection{Factorizing contributions to one-loop $g\to gg$
splitting amplitudes}

The result of the triangle and bubble graphs is (after stripping the
coupling and color factors)~\cite{bc}
\begin{equation}
{\cal D}_{g\to g_1g_2}^{\mu,\,\fact} =
  {i\over\sqrt2}{\tau_\Gamma\over3}
    \left(1-\eps\delta_R-{n_f\over N_c}+{n_s\over N_c}\right)(k_1-k_2)^\mu
     \left[\dprod{\pol_1}{\pol_2}-{\dprod{k_1}{\pol_2}
     \dprod{k_2}{\pol_1}\over\dprod{k_1}{k_2}}\right] \,,
\label{GGG}
\end{equation}
where 
\begin{equation}
\tau_\Gamma^{\vphantom{A}} \equiv
  \left({\mu^2\over-s_{12}}\right)^{\eps}c_\Gamma
  {6\over(1-2\eps)(2-2\eps)(3-2\eps)}\,,
\label{taudef}
\end{equation}
and
\begin{equation}
c_\Gamma = {1\over \left(4\pi\right)^{2-\eps}} 
{\Gamma(1+\eps)\Gamma^2(1-\eps)\over\Gamma(1-2\eps)}\,.
\label{cgdef}
\end{equation}
In background field Feynman gauge, which is a convenient 
gauge for evaluating the diagrams, 
${\cal D}_{g\to g_1g_2}^{\mu,\,\fact}$ is finite due to 
a Ward identity, so no discontinuity functions need to be
subtracted off.  The factorizing part of the one loop splitting function,
${\Split}^{\fact}_{-\lambda}(1^{\lambda_1},2^{\lambda_2})$ (where
the $\lambda$'s label helicities and are defined as if all particles
exit the diagram), is derived from ${\cal D}$ by
\begin{equation}
\eqalign{
  A_n^{\fact} & \iscol{1}{2}
  {\cal D}^{\mu,\,\fact}_{g\to g_1g_2}({\lambda_1},{\lambda_2})
    \left(i\sum_{\lambda=\pm}{\pol^{-\lambda}_\mu(K)\pol^{\lambda}_\nu(K)
    \over2 k_1\cdot k_2}\right)
    {\partial\over\partial\pol^\lambda_\nu(K)}
    A^{\tree}_{n-1}(\dots K^\lambda\dots) \cr
  & = \sum_{\lambda=\pm}{\Split}^{\fact}_{-\lambda}
    (1^{\lambda_1},2^{\lambda_2})A^{\tree}_{n-1}(\dots
    K^\lambda\dots).\cr}
\end{equation}
This yields
\begin{equation}
\eqalign{
{\Split}^{\fact}_{-\lambda}(1^{\lambda_1},2^{\lambda_2}) &= 
   -{\tau_\Gamma\over3}\left(1-\eps\delta_R-{n_f\over N_c}
   +{n_s\over N_c}\right){\pol^{-\lambda}(K)
    \!\cdot\!\left(k_1-k_2\right)\over2\sqrt{2}
   \dprod{k_1}{k_2}}\left[\dprod{\pol_1}{\pol_2}-{\dprod{k_1}{\pol_2}
   \dprod{k_2}{\pol_1}\over\dprod{k_1}{k_2}}\right] \,,\cr
 }
\label{DSplit}
\end{equation}
where $K = k_1+k_2$.  Expressed in terms of spinor inner
products~\cite{SpinorHelicity,mpReview} the result is that
\begin{equation}
\eqalign{
  {\Split}^{\fact}_{+}(1^{+},2^{+}) =\ & 
  -{\tau_\Gamma\over3}\left(1-\eps\delta_R
    -{n_f\over N_c}+{n_s\over N_c}\right)
    \sqrt{z(1-z)}{\spb1.2\over{\spa1.2}^2}\cr
  {\Split}^{\fact}_{+}(1^{-},2^{-})=\ 
  &-{\tau_\Gamma\over3}\left(1-\eps\delta_R
    -{n_f\over N_c}+{n_s\over N_c}\right)
    \sqrt{z(1-z)}{1\over\spb1.2},\cr
  {\Split}^{\fact}_{+}(1^{+},2^{-})=\ 
  &{\Split}^{\fact}_{+}(1^{-},2^{+})=\ 
  0 \,,\cr}
\label{SplitGGG}
\end{equation}
with the remaining terms, ${\Split}_{-}(1^{\lambda_1},
2^{\lambda_2})$, given by parity inversion.  Except for
${\Split}_{+}(1^{+},2^{+})$ and ${\Split}_{-}(1^{-},2^{-})$ which
vanish at tree level, the factorizing contributions to the one loop
splitting amplitudes are proportional to the tree-level splitting
amplitudes,
\begin{equation}
{\Split}^{\fact}_{-\lambda}(1^{\lambda_1},
  2^{\lambda_2}) = c_\Gamma
  {\Split}^{\tree}_{
  -\lambda}(1^{\lambda_1},2^{\lambda_2})
  r^{g\to g_1g_2,\,{\fact}}_S(-\lambda,1^{\lambda_1},2^{\lambda_2}) \,,
\label{splggg}
\end{equation}
where
\begin{equation}
\eqalign{
  r^{g\to g_1g_2,\,{\fact}}_S(\lambda,1^{\pm},2^{\mp}) & = 
  0 \,, \cr
  r^{g\to g_1g_2,\,{\fact}}_S(\pm,1^{\mp},2^{\mp}) & = 
  \left({\mu^2\over-s_{12}}\right)^\eps
    \left(1-\eps\delta_R-{n_f\over N_c}+{n_s\over N_c}\right){2z(1-z)\over
    (1-2\eps)(2-2\eps)(3-2\eps)}\,.\cr}
\label{rsggg}
\end{equation}

\subsubsection{Factorizing contributions to one-loop $g\to
\overline{q}q$ splitting amplitudes}
For $g\to \overline{q}q$ splitting, the sum of the
color and coupling stripped triangle and bubble graphs is
\begin{equation}
\eqalign{ \widetilde{\cal D}^{\mu,\,\fact}_{g\to\overline{q}_1q_2} & =
   i{c_\Gamma\over\sqrt2}\left({\mu^2\over-s_{12}}\right)^\eps
    \overline{u}_{2}\gamma^\mu v_{1}\left\{
    \vphantom{\left[{(\delta_R)\over(3)}\right.}\right.\cr
  &\left[{13\over6\eps(1-2\eps)} + {1\over3(1-2\eps)(3-2\eps)}
    + {(1-\delta_R)\over(1-2\eps)(2-2\eps)(3-2\eps)}\right]\cr
  &-{1\over N_c^2}\left[-{1\over\eps^2}-{3\over2\eps(1-2\eps)}
    -{1\over1-2\eps}+{1-\delta_R\over(2-2\eps)}
    \right]\cr
  &\left. \null+{1\over N_c}\left[-{2n_f+n_s\over3\eps(1-2\eps)}
    +{2(n_f-n_s)\over3(1-2\eps)(3-2\eps)}\right]
    \vphantom{\left[{(\delta_R)\over(3)}\right.}\right\} \,.\cr}
\label{GQQ}
\end{equation}
Here we see a number of terms that are proportional to the
discontinuity functions $f_1$ and $f_2$.  After removing these terms,
we have,
\begin{equation}
\eqalign{
  {\cal D}^{\mu,\,\fact}_{g\to\overline{q}_1q_2} & =
   i{c_\Gamma\over\sqrt2}\left({\mu^2\over-s_{12}}\right)^\eps
    \overline{u}_{2}\gamma^\mu v_{1}\left\{
    {5-2\eps-3\delta_R\over3(1-2\eps)(2-2\eps)(3-2\eps)}\right.\cr
  &\left.\null +{1\over N_c^2}{1+\delta_R(1-2\eps)\over(1-2\eps)(2-2\eps)}
    +{n_f-n_s\over N_c}{2\over3(1-2\eps)(3-2\eps)}\right\}\,. \cr}
\label{GQQa}
\end{equation}

After inserting a complete set of helicity states in an expression
analogous to \eqn{DSplit}, we find that the one loop splitting
function is proportional to the tree-level result for each allowed
helicity configuration.  With $r^{g\to\overline{q}_1q_2,\,{\fact}}_S$
defined as in \eqn{splggg}, we have,
\begin{equation}
\eqalign{
 r^{g\to\overline{q}_1q_2,\,{\fact}}_S(\lambda,1^\pm,2^\mp)
    & = \left({\mu^2\over-s_{12}}\right)^\eps
    \left[{5-2\eps-3\delta_R\over3(1-2\eps)(2-2\eps)(3-2\eps)}
    +{1\over N_c^2}{1+\delta_R(1-2\eps)\over(1-2\eps)(2-2\eps)}\right.\cr
& \phantom{\left({\mu^2\over-s_{12}}\right)^\eps\qquad}
\left. \null +{n_f-n_s\over N_c}{2\over3(1-2\eps)(3-2\eps)}\right] \,. \cr}
\end{equation}

\subsubsection{Factorizing contributions to one-loop $q\to
qg$ splitting amplitudes}
For $q\to qg$ splitting, the sum of the coupling and color
stripped triangle and bubble graphs is
\begin{equation}
\eqalign{
  \widetilde{\cal D}^{i,\,\fact}_{q\to q_1g_2} & = 
   -i{c_{\Gamma}\over\sqrt2}
    \left({\mu^2\over-s_{12}}\right)^\eps
    \left[\overline{u}_{i1}\feynsl{\pol}_2-{\overline{u}_{i1}
    \feynsl{k}_2\dprod{k_1}{\pol_2}\over\dprod{k_1}{k_2}}\right]\times\cr
  &\hskip.5in\left[{1\over\eps^2}
    +{N_c^2-1\over N_c^2}{1\over\eps(1-2\eps)}
    -{N_c^2+1\over N_c^2}{1-\eps\delta_R\over(1-2\eps)
           (2-2\eps)}\right] \,.\cr}
\label{QQG}
\end{equation}
After subtracting out terms proportional to the discontinuity
functions, we get
\begin{equation}
  {\cal D}^{i,\,\fact}_{q\to q_1g_2} = 
  i{c_{\Gamma}\over\sqrt2}
    \left({\mu^2\over-s_{12}}\right)^\eps
    \left[\overline{u}_{i1}\feynsl{\pol}_2-{\overline{u}_{i1}
    \feynsl{k}_2\dprod{k_1}{\pol_2}\over\dprod{k_1}{k_2}}\right]
    {N_c^2+1\over N_c^2}{1-\eps\delta_R\over(1-2\eps)(2-2\eps)}\,.
\label{QQGa}
\end{equation}
Following the same procedure as above, we find that
\begin{equation}
\eqalign{
  r^{q\to q_1g_2,\,{\fact}}_S(\pm,1^{\mp},2^\pm) & = 0 \,, \cr
  r^{q\to q_1g_2,\,{\fact}}_S(\pm,1^{\mp},2^{\mp}) & = 
   \left({\mu^2\over-s_{12}}\right)^\eps\left(1+{1\over N_c^2}\right)
    {(1-z)(1-\eps\delta_R)\over(1-2\eps)(2-2\eps)}\,.\cr }
\label{QQGb}
\end{equation}
For $\qb\to g_1\qb_2$ splitting, we need simply interchange $z$ and
$1-z$.

\subsection{Non-factorizing contributions}
\label{sec:nonfact}

In ref.~\cite{bc} it was shown that all non-factorizing
contributions may be linked to infrared divergences which have a
universal structure for an arbitrary number of external legs.  The
coefficients of the infrared divergences may be used to fix the
coefficients of all integral functions associated with non-factorizing
contributions to the splitting amplitudes.  

The integral functions to be used are selected by demanding that
they contain no infrared divergences other than ones which may
appear in the splitting amplitudes.  The infrared divergences of the
integral functions may be obtained from the explicit forms of the
integrals contained in, for example, refs.~\cite{bc,IntegralLong}.  By
systematically stepping through the list of all integrals and their
discontinuities one can construct a list of infrared divergent
functions (and their associated higher order in $\eps$ parts) that may
appear in the soft or collinear splitting amplitudes. 

\begin{figure}[ht]
\centerline{\epsfxsize 2.8 truein \epsfbox{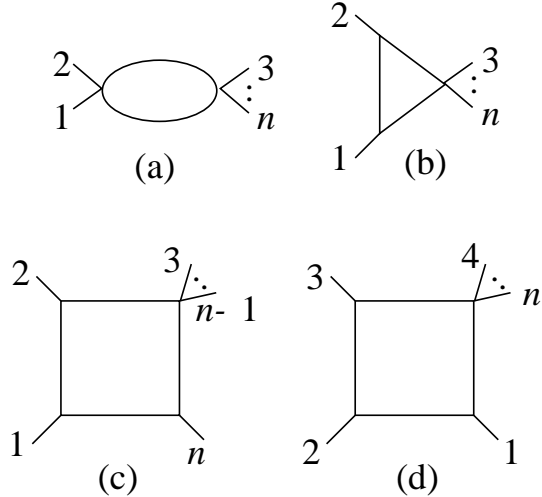}}
\vskip -.2 cm
\caption[]{
\label{IntegralsFigure}
\small  The scalar integrals that can 
appear in the collinear splitting amplitudes.  Legs 1 and 2 
are collinear. }
\end{figure}

More explicitly, the complete one-loop splitting amplitudes are given
by  
\begin{equation}
\Split^{\oneloop} (1, 2) = \Split^{\fact}(1, 2) 
+ c_1 f_1(s_{12}, \eps)
+ c_2 f_2(s_{12}, \eps)
+ c_3 f_3(s_{12}, \eps, z) 
+ c_4 f_4(s_{12}, \eps, z) \,,
\label{GenericSplitLoop}
\end{equation}
where the $c_i$ are coefficients to be fixed using known infrared 
divergences and the discontinuity functions are (see \app{app:BoxApp})
\begin{equation}
\eqalign{
 f_1(s_{12}, \eps) &= - i\mu^{2\eps} {\cal I}_2(s_{12}) = {c_\Gamma
 \over \eps (1-2\eps) } \Bigl({\mu^2 \over-s_{12}} \Bigr)^\eps\,, \cr
 f_2(s_{12}, \eps) &= - i\mu^{2\eps} {\cal I}_3(s_{12}) = - {c_\Gamma
 \over \eps^2} \Bigl( {\mu^2 \over-s_{12}} \Bigr)^\eps \,, \cr
 f_3(s_{12}, \eps, z) &=  -f_2(s_{12}, \eps) + \lim_{k_1 \parallel k_{2}} 
 {i\over 2} \mu^{2\eps}  s_{n 1} s_{12}\, {\cal I}^{\rm 1m}_{4:3} 
= {c_\Gamma \over \eps^2}\left({\mu^2 \over-s_{12}} 
  \right)^\eps\sum_{m=1}^\infty\eps^m{\Li}_m\left({1-z\over-z}
  \right) \,, \cr
f_4(s_{12}, \eps, z) & = -f_2(s_{12}, \eps) + \lim_{k_1 \parallel k_{2}} 
 {i\over 2} \mu^{2\eps}  s_{12} s_{23}\, {\cal I}^{\rm 1m}_{4:4}
 = {c_\Gamma \over
 \eps^2}\left({\mu^2 \over-s_{12}} 
  \right)^\eps\sum_{m=1}^\infty\eps^m{\Li}_m\left({-z\over1-z}
  \right) \,, \cr}
\label{ffunctions}
\end{equation}
where ${\Li}_m(x)$ is the $m{\rm th}$ polylogarithm~\cite{lewin}
\begin{equation}
\left. \begin{array}{l}
{\Li}_1(x)\ = - \ln(1-x) \\ \displaystyle
{\Li}_m(x) = \int_0^x {dt\over t}\,{\Li}_{m-1}(t) \qquad (m=2,3,\dots)
\end{array}\right\} = \sum_{n=1}^{\infty} {x^n\over n^m},\nonumber
\label{polylogdef}
\end{equation}
The scalar integrals appearing in the discontinuity functions
are 
\begin{equation}
\eqalign{
{\rm (a)}\hskip .5 cm &  
{\cal I}_2(s_{12})  = 
\int {d^{4-2\eps} p \over (2\pi)^{4-2\eps} }\,  
{1\over p^2 (p-k_1-k_2)^2} \,,\cr
{\rm (b)}\hskip .5 cm &
{\cal I}_3(s_{12})  = 
\int {d^{4-2\eps} p \over (2\pi)^{4-2\eps} }\,  
{1\over p^2 (p-k_1)^2 (p-k_1-k_2)^2} \,,\cr
{\rm (c)}\hskip .5 cm &
{\cal I}^{\rm 1m}_{4:3}(s_{n1}, s_{12}) = 
\int {d^{4-2\eps} p \over (2\pi)^{4-2\eps} }\, 
{1\over p^2 (p-k_1)^2 (p-k_1-k_2)^2 (p + k_n)^2}\,, \cr
{\rm (d)}\hskip .5 cm &
{\cal I}^{\rm 1m}_{4:4}(s_{12}, s_{23})  = 
\int {d^{4-2\eps} p \over (2\pi)^{4-2\eps} }\, 
{1\over p^2 (p-k_1)^2 (p-k_1-k_2)^2 (p - k_1 - k_2 - k_3)^2} \,. \cr}
\label{LoopMomentumRep}
\end{equation}
These integrals are depicted in \fig{IntegralsFigure}.
No other integrals appear because they are either smooth functions in
the collinear limits, or they do not have the correct infrared
divergences to match those appearing in the splitting
amplitudes~\cite{bc}.  It is easy to see from \eqn{ffunctions} that
$f_4(s_{12}, \eps, z) = f_3(s_{12}, \eps, 1-z)$.  This relation is
apparent even in the integral function by shifting the loop momentum
of ${\cal I}^{\rm 1m}_{4:4}(s_{12}, s_{23})$ such that
$p\to-p+k_1+k_2$.

To fix the coefficients $c_i$ we proceed as follows.  An $n$-point 
primitive amplitude will have a known singular structure of the 
form~\cite{ks,gg,kstsingular}, 
\begin{equation}
A_{n}^{\oneloop} \Bigr|_{\rm singular} = 
 -c_\Gamma A_n^{\tree}\left[{1\over\eps^2}
  \sum_{j=1}^{n}s_j^{[n]} \left({\mu^2\over-s_{j,j+1}}\right)^\eps
  + {c^{[n]}\over\eps}  \right]\,,
\end{equation}
where the coefficients $s^{[n]}_j$ and $c^{[n]}$ depend on the
particular $n$-point amplitude under consideration.  Consider now the
collinear or soft limits given in
\eqns{OneLoopCollinear}{OneLoopSoft}.  The $n-1$ point amplitudes
appearing on the right-hand-side of these equations can have a
different set of singularities.  The difference between these two sets
of known singularities must then be absorbed into the splitting or
soft amplitudes $\Split^\oneloop$ or $\Soft^\oneloop$.  More directly, we
can use the known results for the splitting amplitudes through
$\Ord(\eps^0)$, which we have collected in \app{app:LoopSplitAppendix}, 
to determine the singularities.
Using \tab{NonFactorizingTable} we can then uniquely determine all
$c_i$ coefficients appearing in \eqn{GenericSplitLoop}.  The
unique $1/\eps$ singularities of the functions are listed in the first
column of the table.  The second column lists the discontinuity
function associated with each singularity. One adjusts the $c_i$
coefficients so that the divergences in the splitting amplitudes are
correct.  This uniquely fixes the splitting amplitudes to all
orders in the dimensional regularization parameter $\eps$.

\def\disp{\displaystyle}
\begin{table}[ht]
\begin{center}
\begin{tabular}{|c|c|}
\hline
 {\bf Singularity } 
    &{\bf Non-Factorizing Contribution } \\
\hline
       $ \disp \vphantom{\Bigg|^A} {c_\Gamma\over \eps}$ 
     & $ \disp f_1(s_{12}, \eps)$ \\
\hline
       $  \disp\vphantom{\Bigg|^A} {c_\Gamma\ln(-s_{i, i+1}) \over \eps} $
     & $\disp f_2(s_{12}, \eps)$ \\
\hline
       $ \disp \vphantom{\Bigg|^A} {c_\Gamma\ln(z) \over \eps} $
     & $\disp f_3(s_{12}, \eps, z) $ \\
\hline
       $\disp \vphantom{\Bigg|^A} {c_\Gamma\ln(1-z) \over \eps} $
     & $\disp  f_3(s_{12}, \eps, 1-z) $  \\
\hline
\end{tabular}
\end{center}
\caption[]{
\label{NonFactorizingTable}
\small The potential `non-factorizing' contributions to the
splitting amplitudes in the $k_1 \parallel k_2$ channel.
\smallskip}
\end{table}

Since the divergences of the primitive amplitudes are always 
proportional to tree amplitudes, the non-factorizing
contributions to one-loop splitting amplitudes will
all be proportional to the tree-level splitting amplitudes.  In analogy
to \eqn{splggg}, we write
\begin{equation}
{\Split}^{\nonfact}_{-\lambda}(1^{\lambda_1},
  2^{\lambda_2}) = c_\Gamma \, 
  {\Split}^{\tree}_{
  -\lambda}(1^{\lambda_1},2^{\lambda_2})\,
  r^{\nonfact}_S\,,
\label{rsnonfact}
\end{equation}
for each type of splitting function.  There is no need to specify the
helicity structure of $r^{\nonfact}_S$ because the factor is the same
for all allowed helicity configurations.  For the ${\cal D}$ functions
we can write
\begin{equation}
{\cal D}^{\mu,\,\nonfact}_{K\to1,2} = c_\Gamma \,{\cal
  D}^{\mu,\,\tree}_{K\to1,2} \,r^{\nonfact}_S\,.
\label{Drsnonfact}
\end{equation}

\subsubsection{Non-factorizing contributions to one loop
 $g\to gg$ splitting amplitudes}

In order to demonstrate the procedure more explicitly, we 
extract the non-factorizing contributions to one-loop
$g\to gg$ splitting using pure gluon amplitudes.  The
unrenormalized pole structure of the leading $n$ gluon partial
amplitude $A_{n:1}(1,2,\dots,n)$ is 
\begin{equation}
A^{ng}_{n:1}\Big|_{\rm singular} =\
  -c_\Gamma A_n^{ng,\,\tree}\left[{1\over\eps^2}
  \sum_{j=1}^{n}\left({\mu^2\over-s_{j,j+1}}\right)^\eps
  + {2\over\eps}\left({11\over6}-{1\over3}{n_f\over N_c}
  - {1\over6}{n_s\over N_c}\right)\right]\,,
\label{leadgluepoles}
\end{equation}
where we identify the label $n+1$ with $1$.  In the limit that gluons
$1$ and $2$ become collinear, we verify that for the helicity
configurations forbidden at tree-level,
\begin{equation}
\Split^{g\to g_1g_2,\,{\nonfact}}(\pm,1^\pm,2^\pm) = 0\,,
\label{rsglupolea}
\end{equation}
while for all configurations allowed at tree-level
\begin{equation}
r^{g\to g_1g_2,\,{\nonfact}}_S =  
  -{1\over\eps^2} + {\ln(-s_{12})\over\eps}
  + {\ln(z)\over\eps} + {\ln(1-z)\over\eps} + {\cal O}(\eps^0)\,.
\label{rsglupoleb}
\end{equation}
This may also be read off from the known
results~\cite{bddkSusy4,bdk3g2q} for the splitting amplitudes through
$\Ord(\eps^0)$ which we have collected in \app{app:LoopSplitAppendix}.
Focusing on the $1/\eps$ poles we use \tab{NonFactorizingTable} to
deduce that
\begin{equation}
r^{g\to g_1g_2,\,{\nonfact}}_S =\ 
  f_2(s_{12},\eps) + f_3(s_{12},\eps,z) + f_3(s_{12},\eps,1-z)\,,
\label{rsgluons}
\end{equation}
where the discontinuity functions $f_i$ are defined in
\eqn{ffunctions}.  The combination of functions appearing here may be
further simplified using \eqn{nonfactall}.

\subsubsection{Non-factorizing contributions to one-loop
 $g\to \overline{q}q$ splitting amplitudes}
To extract the non-factorizing contributions to one-loop
 $g\to \overline{q}q$ splitting we need the unrenormalized pole
structure of the leading all-gluon partial amplitudes,
\eqn{leadgluepoles}, as well as that of the leading two quark partial
amplitudes $A_{n:1}(1_{\overline{q}},2_q,3,\dots,n)$,
\begin{equation}
A^{2q}_{n:1}\Big|_{\rm singular} =\
  -c_\Gamma A_n^{2q,\,\tree}\left[{1\over\eps^2}
  \sum_{j=2}^{n}\left({\mu^2\over-s_{j,j+1}}\right)^\eps
  - {1\over N_c^2}{1\over\eps^2}\left({\mu^2\over-s_{12}}\right)^\eps 
  + {3\over2\eps}\left(1-{1\over N_c^2}\right)\right] \,.
\label{leadqqpoles}
\end{equation}
From this it follows that
\begin{equation}
\eqalign{
r^{g\to \overline{q}_1q_2,\,{\nonfact}}_S =&\cr
    {1\over\eps^2}{1\over N_c^2} 
   + {1\over\eps}\left(\ln(z) +\vphantom{{\ln(-s_{12})\over N_c^2}}\right. &
   \left. \ln(1-z) - {\ln(-s_{12})\over N_c^2}
   +{13\over6}+{3\over2N_c^2}-{2n_f\over3}
   -{n_s\over3}\right)+{\cal O}(\eps^0)\,. \cr 
}
\label{gqqnonfac}
\end{equation}
As before, this may also be obtained from the known results for the 
splitting amplitudes through $\Ord(\eps^0)$ which we have collected
in \app{app:LoopSplitAppendix}.
Again, focusing on the $1/\eps$ poles, we read off
\begin{equation}
\eqalign{
r^{g\to \overline{q}_1q_2,\,{\nonfact}}_S =\ 
   f_3(s_{12},\eps,z)& + f_3(s_{12},\eps,1-z)
   - {1\over N_c^2}f_2(s_{12},\eps)\cr  
 & + \left[{13\over6} + {3\over2N_c^2} -{2n_f\over3N_c} -
   {n_s\over3N_c}\right]f_1(s_{12},\eps)\,.\cr 
}
\label{gqqnonfacprim}
\end{equation}

\subsubsection{Non-factorizing contributions to one loop
 $q\to qg$ splitting amplitudes}
We need only the pole structure of the leading two quark partial
amplitudes, \eqn{leadqqpoles}, to derive the non-factorizing
contributions to one-loop $q\to qg$ splitting.  From
\begin{equation}
r^{q\to q_1g_2,\,{\nonfact}}_S
 = -{1\over\eps^2}
   + {1\over\eps}\left(\ln(-s_{12}) + \ln(1-z) - {\ln(z)\over
   N_c^2}\right)+{\cal O}(\eps^0)\,,
\label{qqgnonfac}
\end{equation}
we obtain
\begin{equation}
r^{q\to q_1g_2,\,{\nonfact}}_S
 = f_2(s_{12},\eps) + f_3(s_{12},\eps,1-z)
  - {1\over N_c^2}f_3(s_{12},\eps,z)\,. 
\label{qqgnonfact}
\end{equation}
Again, for $\qb\to g_1\qb_2$ splitting we simply interchange $z$ and
$1-z$.

\subsection{Full one-loop splitting amplitudes for leading color partial
amplitudes to all orders in $\eps$.}
\label{sec:LPAsplit} 
We now combine the factorizing and non-factorizing terms to give the
full one loop splitting amplitudes for leading partial amplitudes to
all orders in $\eps$.

For $g\to g_1g_2$ splitting, there is one independent 
helicity configuration that
is non-zero at one loop but vanishes at tree-level,
\begin{equation}
\eqalign{
\Split^{g\to gg,\,{\oneloop}}_{+}(1^{+},2^{+}) =\ &\cr
     -\sqrt{z(1-z)}{\spb1.2\over{\spa1.2}^2}
  &{2c_\Gamma\over(1-2\eps)(2-2\eps)(3-2\eps)}
     \left({\mu^2\over-s_{12}}\right)^{\eps}
     \left(1-\eps\delta_R -{n_f\over N_c}+{n_s\over N_c}\right)\,.\cr}
\label{fullggga}
\end{equation}
We present the rest of the splitting amplitudes in terms of $r_S$.
\begin{equation}
\eqalign{
r^{g\to g_1g_2,\,{\oneloop}}_{S}(\pm,1^\mp,2^\mp) =\ 
   \left({\mu^2\over-s_{12}}\right)^{\eps}
   &\left\{{1\over\eps^2}\left[-\left({1-z\over z}\right)^\eps 
    {\pi\eps\over \sin(\pi\eps)} + \sum_{m=1}^\infty 2\eps^{2m-1} 
    {\Li}_{2m-1}\left({-z\over 1-z}\right)\right]\right.\cr
   &\left. \null + {2z(1-z)\over(1-2\eps)(2-2\eps)(3-2\eps)}
    \left(1-\eps\delta_R -{n_f\over N_c}+{n_s\over N_c}\right)
    \vphantom{\left[\sum_{m=0}^\infty\right]}\right\}\,,\cr
r^{g\to g_1g_2,\,{\oneloop}}_{S}(\lambda,1^\pm,2^\mp) =\ 
   \left({\mu^2\over-s_{12}}\right)^{\eps}&
    {1\over\eps^2}\left[-\left({1-z\over z}\right)^\eps 
    {\pi\eps\over \sin(\pi\eps)} + \sum_{m=1}^\infty 2\eps^{2m-1} 
    {\Li}_{2m-1}\left({-z\over 1-z}\right)\right]\,.\cr
}
\label{fullgggb}
\end{equation}
In terms of formal polarization vectors,
\begin{equation}
\eqalign{
{\cal D}^{\mu,\,\oneloop}_{g\to g_1g_2} =\ &
    {i\sqrt2}(k_1-k_2)^\mu\left[
    \dprod{\pol_1}{\pol_2}-{\dprod{k_1}{\pol_2}
    \dprod{k_2}{\pol_1}\over\dprod{k_1}{k_2}}\right]
    \left({\mu^2\over-s_{12}}\right)^{\eps}{c_\Gamma
    \left(1-\eps\delta_R-{n_f\over N_c}+{n_s\over N_c}\right)
    \over(1-2\eps)(2-2\eps)(3-2\eps)}\cr
  & + i\sqrt{2} [k_1^\mu \pol_1\cdot \pol_2 
  + k_2\cdot\pol_1 \pol_2^\mu - (k_1 + k_2) \cdot \pol_2 \pol_1^\mu]\times\cr
   &\hskip 1cm
    \left({\mu^2\over-s_{12}}\right)^{\eps}
    {c_\Gamma\over\eps^2}\left[-\left({1-z\over z}\right)^\eps 
    {\pi\eps\over \sin(\pi\eps)} + \sum_{m=1}^\infty 2\eps^{2m-1} 
    {\Li}_{2m-1}\left({-z\over 1-z}\right)\right]\,.\cr}
\label{Formalfullggg}
\end{equation}

For $g\to\qb_1q_2$ splitting, we obtain the same result for each
allowed helicity configuration
\begin{equation}
\eqalign{
r^{g\to\qb_1q_2,\,{\oneloop}}_{S}(\lambda,1^\pm,2^\mp) =\  
   \left({\mu^2\over-s_{12}}\right)^\eps&\left\{
    {1\over\eps^2}\left[1-\left({1-z\over z}\right)^{\eps}
    {\pi\eps\over\sin(\pi\eps)}+\sum_{m=1}^\infty 2\eps^{2m-1} 
    {\Li}_{2m-1}\left({-z\over 1-z}\right)\right]
      \vphantom{\left({\mu^2\over-s_{12}}\right)^\eps}\right.\cr
   &+{13\over6}{1\over\eps(1-2\eps)} + {5-2\eps-3\delta_R\over
     3(1-2\eps)(2-2\eps)(3-2\eps)}
      \vphantom{\left({\mu^2\over-s_{12}}\right)^\eps}\cr
   &+{1\over N_c^2}\left[{1\over\eps^2} + {3\over2}{1\over\eps(1-2\eps)}
      + {1+\delta_R(1-2\eps)\over(1-2\eps)(2-2\eps)}\right]
      \vphantom{\left({\mu^2\over-s_{12}}\right)^\eps}\cr
   &+{n_f\over N_c}\left[-{2\over3}{1\over\eps(1-2\eps)}
      + {2\over3(1-2\eps)(3-2\eps)}\right]
      \vphantom{\left({\mu^2\over-s_{12}}\right)^\eps}\cr
   &+\left.{n_s\over N_c}\left[-{1\over3}{1\over\eps(1-2\eps)}
      - {2\over3(1-2\eps)(3-2\eps)}\right]
   \vphantom{\left({\mu^2\over-s_{12}}\right)^\eps}\right\} \,.\cr
}
\label{fullgqq}
\end{equation}
For $g\to q_1\qb_2$ splitting we simply interchange $z$ and $1-z$.
In fact, the $g\rightarrow \bar q q$ and $g\rightarrow gg$ splitting 
functions are symmetric in $z\rightarrow1-z$, as can be seen from the
results of \app{app:BoxApp}.
In terms of formal polarization vectors the ${\cal D}$ function is
\begin{equation}
{\cal D}^{\mu,\,\oneloop}_{g\to\qb_1q_2} = {i \over \sqrt{2} }
  \overline{u}_2 \gamma^\mu v_1 c_\Gamma \, 
r^{g\to\qb_1q_2,\,{\oneloop}}_{S}(\lambda,1^\pm,2^\mp) 
\label{Formalfullgqq}
\end{equation}

Finally, for $q\to q_1g_2$ and $\qb\to g_1\qb_2$ splitting,
\begin{equation}
\eqalign{
r^{q\to q_1g_2,\,{\oneloop}}_S(\pm,1^{\mp},2^{\mp}) =\ 
   &-\left(\mu^2\over-s_{12}\right)^\eps\left\{{1\over\eps^2}
    \left[1-\sum_{m=1}^{\infty}\eps^m\left[
    \Li_m\left({-z\over1-z}\right) - {1\over N_c^2}\Li_m
    \left({1-z\over-z}\right)\right]\right]\right.\cr
   &\left.\hskip 2.5cm\vphantom{\sum_{m=1}^{\infty}}
    - \left(1+{1\over N_c^2}\right){(1-z)(1-\eps\delta_R)
    \over(1-2\eps)(2-2\eps)}\right\}\,,\cr
r^{q\to q_1g_2,\,{\oneloop}}_S(\pm,1^{\mp},2^\pm) =\ 
   &-\left(\mu^2\over-s_{12}\right)^\eps{1\over\eps^2}
   \left[1-\sum_{m=1}^{\infty}\eps^m \left[
   \Li_m\left({-z\over1-z}\right) - {1\over N_c^2}\Li_m
   \left({1-z\over-z}\right)\right]\right]\,,\cr
r^{\qb\to g_1\qb_2,\,{\oneloop}}_S(\pm,1^{\mp},2^{\mp}) =\ 
   &-\left(\mu^2\over-s_{12}\right)^\eps\left\{{1\over\eps^2}
   \left[1-\sum_{m=1}^{\infty}\eps^m\left[
    \Li_m\left({1-z\over-z}\right) - {1\over N_c^2}\Li_m
    \left({-z\over1-z}\right)\right]\right]\right.\cr
   &\left.\hskip 2.5cm\vphantom{\sum_{m=1}^{\infty}}
    - \left(1+{1\over N_c^2}\right)
     {z(1-\eps\delta_R)\over(1-2\eps)(2-2\eps)}\right\}\,,\cr
r^{\qb\to g_1\qb_2,\,{\oneloop}}_S(\pm,1^{\mp},2^\pm) =\ 
   &-\left(\mu^2\over-s_{12}\right)^\eps{1\over\eps^2}
    \left[1-\sum_{m=1}^{\infty}\eps^m\left[
    \Li_m\left({1-z\over-z}\right) - {1\over N_c^2}\Li_m
    \left({-z\over1-z}\right)\right]\right]\,.\cr
}
\label{fullqqg}
\end{equation}
For $q\to g_1q_2$ and $\qb\to \qb_1g_2$ splitting we simply
interchange $z$ and $1-z$.  
In terms of formal polarization vectors, these are written as
\begin{equation}
\eqalign{
{\cal D}^{i,\,\oneloop}_{q\to q_1g_2} =\ &
    {i\over\sqrt2}\left[\overline{u}_{i1}\feynsl{\pol}_2
    -{\overline{u}_{i1}\feynsl{p}_2\dprod{k_1}{\pol_2}\over
    \dprod{k_1}{k_2}}\right]c_{\Gamma} \left({\mu^2\over
    -s_{12}}\right)^\eps\left(1+{1\over N_c^2}\right)
    {1-\eps\delta_R\over(1-2\eps)(2-2\eps)}\cr
    & -{i\over\sqrt{2}}\overline{u}_{j1}\feynsl{\pol}_2
    c_\Gamma\left({\mu^2\over-s_{12}}\right)^\eps{1\over\eps^2}
   \left[1-\sum_{m=1}^{\infty}\eps^m \left[
   \Li_m\left({-z\over1-z}\right) - {1\over N_c^2}\Li_m
   \left({1-z\over-z}\right)\right]\right]\,,\cr
}
\label{Formalfullqqg}
\end{equation}

\subsection{Renormalizing the one-loop splitting amplitudes}
\label{BareToRenormalizedSubSubSection}

We can easily convert to the renormalized splitting amplitudes by
performing an $\overline{\rm MS}$ ultraviolet subtraction\footnote{The
$\overline{\rm MS}$ subtraction \cite{bardeen} is precisely defined
only to ${\cal O}(\eps^0)$. Beyond ${\cal O}(\eps^0)$ there are
several different definitions \cite{cat,van,samuel} and \cite{kstfour}
(which contains two different definitions in sect.~4 and 5).  We
choose the second definition of ref.~\cite{kstfour}.} using the fact
that the splitting amplitudes implicitly contain one power of the
coupling\footnote{The $\overline{\rm MS}$ running coupling depends on
the regularization scheme in which the one-loop amplitude is
renormalized; {\it e.g.\/} in the 't Hooft-Veltman scheme, renormalizing the
amplitude through \eqn{nonfactrenorm} amounts to replacing the bare
coupling $\alpha_s$ with the running coupling $\alpha_s^{HV}$.  The
difference between running couplings in the different schemes is of
$O(\epsilon^0)$ \cite{kstfour,cst}.} and that the ultraviolet divergence of an
$n$-point one-loop amplitude is given by
\begin{equation}
A^{\oneloop}_{n} \Bigr|_{\rm UV\ divergence } 
= c_\Gamma A^{\tree}_n{n-2\over \eps}
  \left({11\over6} - {1\over3}{n_{f}\over N_c}
    - {1\over6}{n_{s}\over N_c}\right)\,.  
\label{AnRenorm}
\end{equation}
All other poles in $\eps$ in $A^{\oneloop}_{n}$ are infrared
singularities.  Since all of the singular terms have been pushed into
the non-factorizing term, renormalization can be accomplished by
altering just that piece:
\begin{equation}
r^{a\to b_1c_2,\,{\nonfact}}_{S,\,{\rm ren}} = 
r^{a\to b_1c_2,\,{\nonfact}}_{S} - {1\over \eps} \left({11\over 6}
    -{1\over 3}{n_{f}\over N_c} -{1\over 6}{n_{s}\over N_c}\right) \,.
\label{nonfactrenorm}
\end{equation}
Following this procedure we can rewrite any of the splitting or 
soft amplitudes presented in this paper so that they hold for 
renormalized instead of bare amplitudes.

\subsection{Primitive decomposition of one-loop splitting amplitudes}
\label{sec:PrimSplit}

As discussed in \sec{sec:ColorRev}, one-loop amplitudes are naturally
broken down into color ordered partial amplitudes, which can
themselves be broken down into gauge invariant primitive components.
Once we determine the splitting rules for the primitive amplitudes,
then formulae like \eqn{sublanswer} can be used to generate the
splitting rules for any partial amplitude.

In fact, collinear splitting is a property of full partial amplitudes,
not just of the primitive ones.  Furthermore, it is only the leading
partial amplitude, $A_{5;1}$ which has the same color structure as the
tree-level amplitude, that receives splitting contributions from both
$\Split^{\tree}$ and $\Split^{\oneloop}$.  The sub-leading partial
amplitudes, which have no tree-level counterparts, require only
$\Split^{\tree}$.  The primitive components of the sub-leading partial
amplitudes do receive splitting contributions from
$\Split^{\oneloop}$, but these contributions cancel in the permutation
sums that build the sub-leading partial amplitudes.  This can easily
be seen for the all-gluon case using \eqn{sublanswer},
\begin{equation}
\eqalign{
A_{n;j}(1,2,\ldots,j-1;j,\ldots,n)\ \iscol{1}{2}\ &(-1)^{j-1}
  \Split^{\tree}(1,2)\sum_{\sigma\in COP\{\alpha\}
   \{\beta\}}A^{[1]}_{n-1;1}(\sigma)\cr
 +&(-1)^{j-1}\Split^{\oneloop}(1,2)\sum_{\sigma\in COP\{\alpha\}
   \{\beta\}}A^{\tree}_{n-1}(\sigma)\,,\cr
}
\label{collsublanswer}
\end{equation}
where $\alpha_i \in \{\alpha\} \equiv \{j-1,j-2,\ldots,3,c\}$,
$\beta_i \in \{\beta\} \equiv \{j,j+1,\ldots,n-1,n\}$, and $c$
is the coalescence of gluons $1$ and $2$.  The sum of tree-level
amplitudes over $COP$ permutations vanishes~\cite{bg}.  The same
permutation sums (see \eqn{subl2quark}) and subsequent cancellation of
tree-level amplitudes occurs in the collinear splitting of two-quark
sub-leading partial amplitudes as well. 

The primitive decomposition of the splitting amplitudes is thus not
important to the computation of the one-loop cross section in the
collinear limit.  This should not be too surprising since collinear
factorization is a global property of the entire tree-level amplitude,
not just of the color-ordered amplitudes, which play the role of
tree-level primitive amplitudes.  
Nonetheless, the primitive splitting rules are
important for understanding the structure of primitive amplitudes and
serve as useful checks on the primitive amplitudes themselves.
Moreover, the primitive decomposition is important for an
understanding of universal soft factorization.  Even at tree-level,
soft factorization is only a property of color ordered sub-amplitudes,
not of the amplitude as a whole.  Since soft factorization becomes
apparent at the primitive level, it is convenient to express all
infrared behavior at that level.

We can derive the primitive splitting rules for the all-gluon
and the two-quark amplitudes directly from the splitting
rules of the leading partial amplitude.  The reason for this is the
simple primitive decomposition rule for the leading partial
amplitudes.  For instance, for $n$ gluon amplitudes,
\begin{equation}
A_{n:1}(1,2,\dots,n) = A^{[1]}_{n;1}(1,2,\dots,n)
   + {n_f\over N_c}A^{[1/2]}_{n;1}(1,2,\dots,n)
   + {n_s\over N_c}A^{[0]}_{n;1}(1,2,\dots,n) \,,
\label{Aoneglue}
\end{equation}
where $A^{[J]}_{n;1},\,J=1,1/2,0$ are the primitive amplitudes.
Corresponding to these primitive amplitudes are three one-loop
primitive splitting amplitudes, ${\Split}^{[J]}$.
The leading terms in $N_c$ in the splitting relation for $A_{n;1}$
are part of the primitive splitting function ${\Split}^{[1]}$,
those proportional to $n_f/N_c$ are part of ${\Split}^{[1/2]}$, and 
those proportional to $n_s/N_c$ are part of ${\Split}^{[0]}$.  That
is,
\begin{equation}
A_{n;1}\  \iscol{1}{2}\ {\Split}^{\tree}A_{n-1;1} 
 +{\Split}^{[1]}A^{\tree}_{n-1}
 +{n_f\over N_c}{\Split}^{[1/2]}A^{\tree}_{n-1}
 +{n_s\over N_c}{\Split}^{[0]}A^{\tree}_{n-1} \,,
\label{LeadPartSplit}
\end{equation}
which breaks up into terms like \eqn{OneLoopCollinear},
\begin{equation}
\eqalign{
A^{[1]}_{n;1}\  \iscol{1}{2}\ & {\Split}^{\tree}A^{[1]}_{n-1;1} 
 +{\Split}^{[1]}A^{\tree}_{n-1}\,,\cr
A^{[1/2]}_{n;1}\  \iscol{1}{2}\ & {\Split}^{\tree}A^{[1/2]}_{n-1;1} 
 +{\Split}^{[1/2]}A^{\tree}_{n-1}\,,\cr
A^{[0]}_{n;1}\  \iscol{1}{2}\ & {\Split}^{\tree}A^{[0]}_{n-1;1} 
 +{\Split}^{[0]}A^{\tree}_{n-1}\,,\cr}
\label{LeadPartSplita}
\end{equation}
where the helicity sum has been suppressed.

Similarly, when considering amplitudes with a pair of external
quarks, there are four types of primitive amplitudes, $A^{L,[1]}_n$,
$A^{R,[1]}_n$, $A^{L,[1/2]}_n$ and $A^{L,[0]}_n$ in \eqn{Anoneformula},
and, accordingly, four primitive splitting amplitudes.  The leading
terms in $N_c$ in the splitting relation for $A_{n:1}$ correspond to
${\Split}^L$, those suppressed by $1/N_c^2$ correspond to
${\Split}^R$, {\it etc}.

\subsubsection{Primitive splitting rules for all-gluon amplitudes}
Using \eqn{LeadPartSplita} and the results of \sec{sec:LPAsplit}, the
splitting rules for all-gluon primitive amplitudes are
\begin{eqnarray}
\label{AllGPrimSplitOne}
\eqalign{
\Split^{ng,\,[1]}_{+}(1^{+},2^{+}) =\ 
   &-\left({\mu^2\over-s_{12}}\right)^{\eps}{\spb1.2\over{\spa1.2}^2}
    {2\sqrt{z(1-z)}(1-\eps\delta_R)c_\Gamma\over(1-2\eps)(2-2\eps)(3-2\eps)}
    \,,\cr
r_S^{ng,\,[1]}(\pm,1^\mp,2^\mp) =\ 
   &{1\over\eps^2}\left({\mu^2\over-s_{12}}\right)^{\eps}
    \left[-\left({1-z\over z}\right)^\eps 
    {\pi\eps\over \sin(\pi\eps)} + \sum_{m=1}^\infty 2\eps^{2m-1} 
    {\Li}_{2m-1}\left({-z\over 1-z}\right)\right]\cr
   & + \left({\mu^2\over-s_{12}}\right)^{\eps}
     {2z(1-z) (1- \delta_R \eps) \over(1-2\eps)(2-2\eps)(3-2\eps)}\,,\cr
r_S^{ng,\,[1]}(\lambda,1^\pm,2^\mp) =\ 
   &{1\over\eps^2}\left({\mu^2\over-s_{12}}\right)^{\eps}
    \left[-\left({1-z\over z}\right)^\eps 
    {\pi\eps\over \sin(\pi\eps)} + \sum_{m=1}^\infty 2\eps^{2m-1} 
    {\Li}_{2m-1}\left({-z\over 1-z}\right)\right]\,,\phantom{0}\cr
}\\
\label{AllGPrimSplitHalf}
\eqalign{
\Split^{ng,\,[1/2]}_{+}(1^{+},2^{+}) =\ 
   &+\left({\mu^2\over-s_{12}}\right)^{\eps}{\spb1.2\over{\spa1.2}^2}
    {2\sqrt{z(1-z)} c_\Gamma\over(1-2\eps)(2-2\eps)(3-2\eps)}
    \,,\cr
r_S^{ng,\,[1/2]}(\pm,1^\mp,2^\mp) =\ 
   &-\left({\mu^2\over-s_{12}}\right)^{\eps}
    {2z(1-z)\over(1-2\eps)(2-2\eps)(3-2\eps)}\,,\cr
r_S^{ng,\,[1/2]}(\lambda,1^\pm,2^\mp) =\ &0\,,
    \phantom{{1\over\eps^2}\left({\mu^2\over-s_{12}}\right)^{\eps}
    \left[-\left({1-z\over z}\right)^\eps 
    {\pi\eps\over \sin(\pi\eps)} + \sum_{m=1}^\infty 2\eps^{2m-1} 
    {\Li}_{2m-1}\left({-z\over 1-z}\right)\right]}\cr
}\\
\label{AllGPrimSplitZero}
\eqalign{
\Split^{ng,\,[0]}_{+}(1^{+},2^{+}) =\ 
   &-\left({\mu^2\over-s_{12}}\right)^{\eps}
    {\spb1.2\over{\spa1.2}^2}
    {2\sqrt{z(1-z)} c_\Gamma\over(1-2\eps)(2-2\eps)(3-2\eps)}\,,\cr
r_S^{ng,\,[0]}(\pm,1^\mp,2^\mp) =\ 
   &+\left({\mu^2\over-s_{12}}\right)^{\eps}
    {2z(1-z)\over(1-2\eps)(2-2\eps)(3-2\eps)}\,,\cr
r_S^{ng,\,[0]}(\lambda,1^\pm,2^\mp) =\ &0\,.
    \phantom{{1\over\eps^2}\left({\mu^2\over-s_{12}}\right)^{\eps}
    \left[-\left({1-z\over z}\right)^\eps 
    {\pi\eps\over \sin(\pi\eps)} + \sum_{m=1}^\infty 2\eps^{2m-1} 
    {\Li}_{2m-1}\left({-z\over 1-z}\right)\right]} \,. \cr}
\end{eqnarray}

\subsubsection{Primitive splitting rules for two-quark amplitudes}
\label{sec:Prim2qSplit}
It is clear from \eqn{Anoneformula} that we can obtain splitting rules
for at least some of the primitive amplitudes from the results of
\sec{sec:LPAsplit}.  However, there are additional primitive
amplitudes for two-quark processes, those where the anti-quark and
quark are not adjacent in the ordering like
$A_5^{L,[1]}(1_{\overline{q}}, 3, 2_q, 4, 5)$, that contribute to 
sub-leading, but not the leading,  partial amplitudes.  Nonetheless,
it turns out that all of the rules can be derived from the splitting
of the leading partial amplitudes.

In $g\to gg$ splitting, there are two possibilities: the collinear
pair could lie between the quark and the anti-quark in the ordering,
({\it i.e.\/} on the gluonic part of the loop) as it must in the
leading partial amplitudes, or, as it can in the additional primitive
amplitudes, the collinear pair could lie between the anti-quark and
the quark (on the fermionic part of the loop).  Similarly, the leading
partial amplitudes 
can have $q\to qg$ or $\qb\to g\qb$ splitting, but the additional
primitive amplitudes can also have $q\to gq$ or $\qb\to \qb{g}$ splitting.
Clearly, neither $g\to\qb{q}$ nor $g\to q\qb$ splitting contributes to
the splitting of primitive amplitudes where the quark and anti-quark are not
adjacent in the ordering.

It would seem that we must take these extra cases into account.
However, because the $A_n^{R,\,[J]}$'s are related to the 
$A_n^{L,\,[J]}$'s by a reflection identity, \eqn{LandR}, in which the
ordering is reversed, an `$R$'-type primitive with the collinear
pair lying between the quark and anti-quark is equivalent to an
``$L$''-type primitive with the collinear pair lying between the
anti-quark and the quark.  The `extra' splitting processes are thus
already taken account in the splitting of the leading partial
amplitudes.

We thus need only to derive the splitting rules for the `$L$'-type
primitive amplitudes.  There are eight cases: $g\to gg$ splitting when the
collinear pair lies between the quark and the anti-quark; $g\to gg$
splitting when the collinear pair lies between the anti-quark and the
quark; $q\to qg$ splitting; $q\to gq$ splitting; $\qb\to g\qb$
splitting; $\qb\to\qb{g}$ splitting; $g\to\qb{q}$ splitting; and
$g\to q\qb$ splitting.

For $g\to g_1g_2$ splitting when the collinear pair is between the
quark and anti-quark the splitting functions are identical 
to the pure glue case of eqs.~(\ref{AllGPrimSplitOne}), 
(\ref{AllGPrimSplitHalf}) and (\ref{AllGPrimSplitZero}):
\begin{equation}
\Split^{2q,\,L,\,[J=1,1/2,0]}_{\qb q[g\to gg],\,\lambda}
  (1^{\lambda_1},2^{\lambda_2}) = 
\Split^{ng,\,[J=1,1/2,0]}_{\lambda}
  (1^{\lambda_1},2^{\lambda_2}) \,.
\label{qqPrSplggga}
\end{equation}
For $g\to g_1g_2$ splitting when the collinear pair is between the
anti-quark and quark,
\begin{equation}
\Split^{2q,\,L,\,[J=1,1/2,0]}_{\qb[g\to gg]q,\,\lambda}
  (1^{\lambda_1},2^{\lambda_2}) = 0 \,.
\label{qqPrSplgggb}
\end{equation}
For $g\to\qb_1q_2$ splitting,
\begin{equation}
\eqalign{
r^{2q,\,L,\,[1]}_{S\,[g\to\qb{q}]}(\lambda,1^\pm,2^\mp) =\ 
   \left({\mu^2\over-s_{12}}\right)^\eps&\left\{
    {1\over\eps^2}\left[1-\left({1-z\over z}\right)^{\eps}
    {\pi\eps\over\sin(\pi\eps)}+\sum_{m=1}^\infty 2\eps^{2m-1} 
    {\Li}_{2m-1}\left({-z\over 1-z}\right)\right]
      \vphantom{\left({\mu^2\over-s_{12}}\right)^\eps}\right.\cr
   &+\left.{13\over6}{1\over\eps(1-2\eps)} + {5-2\eps-3\delta_R\over
     3(1-2\eps)(2-2\eps)(3-2\eps)}
      \vphantom{\left({\mu^2\over-s_{12}}\right)^\eps}\right\}\,,\cr
r^{2q,\,L,\,[1/2]}_{S\,[g\to\qb{q}]}(\lambda,1^\pm,2^\mp) =\ 
    \left({\mu^2\over-s_{12}}\right)^\eps
   &\left[-{2\over3}{1\over\eps(1-2\eps)}
      + {2\over3(1-2\eps)(3-2\eps)}\right]\,,\cr
r^{2q,\,L,\,[0]}_{S\,[g\to\qb{q}]}(\lambda,1^\pm,2^\mp) =\ 
     \left({\mu^2\over-s_{12}}\right)^\eps
    &\left[-{1\over3}{1\over\eps(1-2\eps)}
      - {2\over3(1-2\eps)(3-2\eps)}\right]\,.\cr
}
\label{qqPrSplgqqa}
\end{equation}
For $g\to q_1\qb_2$ splitting,
\begin{equation}
\eqalign{
r^{2q,\,L,\,[1]}_{S\,[g\to{q}\qb]}(\lambda,1^\pm,2^\mp) =\ 
   &-\left({\mu^2\over-s_{12}}\right)^\eps
      \left[{1\over\eps^2} + {3\over2}{1\over\eps(1-2\eps)}
      + {1+\delta_R(1-2\eps)\over(1-2\eps)(2-2\eps)}\right]\,,\cr
r^{2q,\,L,\,[1/2]}_{S\,[g\to{q}\qb]}(\lambda,1^\pm,2^\mp) =\ 
   &0\,,\cr
r^{2q,\,L,\,[0]}_{S\,[g\to{q}\qb]}(\lambda,1^\pm,2^\mp) =\ 
   &0\,.\cr}
\label{qqPrSplgqqb}
\end{equation}

In the following $z$ refers to the momentum fraction of parton $1$.
Thus, in $q\to q_1g_2$ splitting, $z$ is the quark momentum fraction,
while in $q\to g_1q_2$ splitting, $z$ is the gluon momentum fraction.
For $q\to q_1g_2$ splitting,
\begin{equation}
\eqalign{
r^{2q,\,L,\,[1]}_{S\,[q\to{q}g]}(\pm,1^\mp,2^\mp) =\ 
   &-\left({\mu^2\over-s_{12}}\right)^\eps\left\{{1\over\eps^2}
    \left[1-\sum_{m=1}^{\infty}\eps^m\Li_m\left({-z\over1-z}\right)
    \right]  - {(1-z)(1-\eps\delta_R)
    \over(1-2\eps)(2-2\eps)}\right\}\,,\cr
r^{2q,\,L,\,[1]}_{S\,[q\to{q}g]}(\pm,1^\mp,2^\pm) =\ 
   &-\left({\mu^2\over-s_{12}}\right)^\eps
      {1\over\eps^2}\left[1-\sum_{m=1}^{\infty}\eps^m
   \Li_m\left({-z\over1-z}\right)\right]\,,\cr
r^{2q,\,L,\,[1/2]}_{S\,[q\to{q}g]}(\lambda,1^{-\lambda},2^\pm) =\ 
   &0\,,\cr
r^{2q,\,L,\,[0]}_{S\,[q\to{q}g]}(\lambda,1^{-\lambda},2^\pm) =\ 
   &0\,.\cr
}
\label{qqPrSplqqg}
\end{equation}
For $q\to g_1q_2$ splitting,
\begin{equation}
\eqalign{
r^{2q,\,L,\,[1]}_{S\,[q\to{g}q]}(\pm,1^\mp,2^\mp) =\ 
   &-\left(\mu^2\over-s_{12}\right)^\eps\left\{{1\over\eps^2}
    \sum_{m=1}^{\infty}\eps^m\Li_m\left({-z\over1-z}\right)
    - {z(1-\eps\delta_R)\over(1-2\eps)(2-2\eps)}\right\}\,,\cr
r^{2q,\,L,\,[1]}_{S\,[q\to{g}q]}(\pm,1^\mp,2^\pm) =\ 
   &-\left(\mu^2\over-s_{12}\right)^\eps{1\over\eps^2}
    \sum_{m=1}^{\infty}\eps^m \Li_m\left({-z\over1-z}\right)\,,\cr
r^{2q,\,L,\,[1/2]}_{S\,[q\to{g}q]}(\lambda,1^{-\lambda},2^\pm) =\ 
   &0\,,\cr
r^{2q,\,L,\,[0]}_{S\,[q\to{g}q]}(\lambda,1^{-\lambda},2^\pm) =\ 
   &0\,.\cr
}
\label{qqPrSplqgq}
\end{equation}
For $\qb\to g_1\qb_2$ splitting,
\begin{equation}
\eqalign{
r^{2q,\,L,\,[1]}_{S\,[\qb\to{g}\qb]}(\pm,1^\mp,2^\mp) =\ 
   &-\left(\mu^2\over-s_{12}\right)^\eps\left\{{1\over\eps^2}
   \left[1-\sum_{m=1}^{\infty}\eps^m\Li_m\left({1-z\over-z}\right)
    \right] - {z(1-\eps\delta_R)\over(1-2\eps)(2-2\eps)}\right\}\,,\cr
r^{2q,\,L,\,[1]}_{S\,[\qb\to{g}\qb]}(\pm,1^\mp,2^\pm) =\ 
   &-\left(\mu^2\over-s_{12}\right)^\eps{1\over\eps^2}
    \left[1-\sum_{m=1}^{\infty}\eps^m\Li_m
    \left({1-z\over-z}\right)\right]\,,\cr
r^{2q,\,L,\,[1/2]}_{S\,[\qb\to{g}\qb]}(\lambda,1^{-\lambda},2^\pm) =\ 
   &0\,,\cr
r^{2q,\,L,\,[0]}_{S\,[\qb\to{g}\qb]}(\lambda,1^{-\lambda},2^\pm) =\ 
   &0\,.\cr
}
\label{qqPrSplqbgqb}
\end{equation}
For $\qb\to\qb_1g_2$ splitting,
\begin{equation}
\eqalign{
r^{2q,\,L,\,[1]}_{S\,[\qb\to\qb{g}]}(\pm,1^\mp,2^\mp) =\ 
   &-\left(\mu^2\over-s_{12}\right)^\eps\left\{{1\over\eps^2}
   \sum_{m=1}^{\infty}\eps^m\Li_m\left({1-z\over-z}\right)
    - {(1-z)(1-\eps\delta_R)\over(1-2\eps)(2-2\eps)}\right\}\,,\cr
r^{2q,\,L,\,[1]}_{S\,[\qb\to\qb{g}]}(\pm,1^\mp,2^\pm) =\ 
   &-\left(\mu^2\over-s_{12}\right)^\eps{1\over\eps^2}
    \sum_{m=1}^{\infty}\eps^m\Li_m\left({1-z\over-z}\right)\,,\cr
r^{2q,\,L,\,[1/2]}_{S\,[\qb\to\qb{g}]}(\lambda,1^{-\lambda},2^\pm) =\ 
   &0\,,\cr
r^{2q,\,L,\,[0]}_{S\,[\qb\to\qb{g}]}(\lambda,1^{-\lambda},2^\pm) =\ 
   &0\,.\cr
}
\label{qqPrSplqbqbg}
\end{equation}

As for the leading color partial amplitudes it is straightforward
to rewrite the primitive splitting amplitudes in terms of formal 
polarization vectors and spinors.

\subsection{One-loop soft amplitudes}

The situation with soft limits is quite similar to the collinear
case.  Again the contributions can be divided into factorizing and
non-factorizing pieces.  The factorizing contributions to the soft
limit of one-loop $n$ point amplitudes come from those diagrams where
an $n-1$ point tree diagram emits a soft gluon via a loop.  These are
exactly the same diagrams that gave the factorizing contribution to
the collinear splitting amplitudes, although we are now taking a
different infrared limit.  Taking the gluon momentum $k_2$ to zero in
\eqns{GGG}{QQGa}, we find that the factorizing contributions to the
one-loop soft amplitudes vanish.

Just as the non-factorizing contributions to collinear splitting came
from taking the limit of infrared singular loop diagrams, the
non-factorizing contributions to the soft amplitudes come from taking
the soft limit of those same loop diagrams.  In
\tab{NonFactorizingSoftTable} we list the  infrared divergences and 
the corresponding discontinuity functions that can appear in the soft
limit.

\begin{table}[ht]
\begin{center}
\begin{tabular}{|c|c|}
\hline
 {\bf Singularity } 
    &{\bf Non-Factorizing Contribution } \\
\hline
       $  \disp\vphantom{\Bigg|^A} {c_\Gamma\ln(-s_{n1}) \over \eps} $
     & $\disp f_2(s_{n1}, \eps)$ \\
\hline
       $  \disp\vphantom{\Bigg|^A} {c_\Gamma \ln(-s_{12}) \over \eps} $
     & $\disp f_2(s_{12}, \eps)$ \\
\hline
       $ \disp \vphantom{\Bigg|^A} {c_\Gamma\ln((-s_{n1}) (-s_{12}) /(-s_{n2})) 
              \over \eps} $
     & $\disp f_5(s_{n1}, s_{12}, s_{n2}, \eps)$ \\
\hline
\end{tabular}
\end{center}
\caption[]{
\label{NonFactorizingSoftTable}
\small The potential `non-factorizing' contributions to the soft amplitudes
for $k_1 \rightarrow 0$.  
\smallskip}
\end{table}

The complete one-loop soft amplitude as the momentum of leg 1 vanishes
is given by
\begin{equation}
\Soft^{\oneloop} (n,1,2) = \Soft^{\fact}(n,1, 2) 
+ s_1 f_2(s_{12}, \eps)
+ s_2 f_2(s_{n1}, \eps)
+ s_3 f_5(s_{n1}, s_{12}, s_{n2}, \eps) \,,
\label{loopsoftfundef}
\end{equation}
where $f_2$ is defined in \eqn{ffunctions} and $f_5$ is (see
\app{app:SoftFun})
\begin{equation}
f_5(s_{n1}, s_{12},  s_{n2}, \eps) \equiv \lim_{k_1 \to 0}
 {i\over 2} \mu^{2\eps} s_{n1} s_{12}\, 
       {\cal I}^{\rm 1m}_{4:3}
 =-{c_{\Gamma}\over\epsilon^2}\,
\left({\mu^2(-s_{n2})\over (-s_{n1})(-s_{12})}\right)^{\epsilon}\, 
{\pi\epsilon\over \sin(\pi\epsilon)} \,.
\label{f5def}
\end{equation}

Since $\Soft^{\fact}(n,1, 2)=0$, all that is needed to get the
complete soft amplitude is to examine the infrared singularities of
one-loop amplitudes and adjust the coefficients $s_i$ so that we have
the correct poles in $\eps$.

Once again we can obtain the primitive soft rules for zero- and
two-quark amplitudes from the soft rules of the leading partial
amplitudes.  Using the known divergence structure of the soft
amplitudes (\ref{LeadingSoft}), we find the infrared divergence
structure,
\begin{equation}
\eqalign{
\Soft^{\nonfact} =& c_\Gamma\Soft^{\tree}\left[-{1\over\eps^2}
   + {\ln(-s_{n1})\over\eps} + {\ln(-s_{12})\over\eps}
   - {\ln(-s_{n2})\over\eps}\right] + {\cal O}(\eps^0)\cr
  =& c_\Gamma\Soft^{\tree}\left[-{1\over\eps^2}
   + {\ln(-s_{n1})(-s_{12})/(-s_{n2})\over\eps}\right]
   + {\cal O}(\eps^0)\,.\cr}
\label{SoftNonFact}
\end{equation}
Using \tab{NonFactorizingSoftTable} to match these divergences
to their proper discontinuity function, it readily follows that
\begin{equation}
\eqalign{
\Soft^{\oneloop} (n,1^\pm,2) =& \Soft^{\tree}(n,1^\pm,2)
 f_5(s_{n1}, s_{12}, s_{n2}, \eps)\cr
 =& -{c_\Gamma\over\eps^2}\Soft^{\tree}(n,1^\pm,2)
 \left({\mu^2(-s_{n2})\over (-s_{n1})(-s_{12})}\right)^{\epsilon}\, 
 {\pi\epsilon\over \sin(\pi\epsilon)}\,,\cr}
\label{SoftGluonExact}
\end{equation}
in agreement with our earlier result~\cite{us}.  Although the helicity
form (\ref{HelSoft}) of $\Soft^{\tree}(n,1^\pm,2)$ is usually the most
convenient representation, within the context of conventional
dimensional regularization it is convenient to express
$\Soft^{\tree}(n,1^\pm,2)$ in terms of the eikonal factor in
\eqn{Eikonal}.

This result does not depend on $\delta_R$, so it is
independent of the variety of dimensional regularization scheme used.
This result is also independent of the particle and helicity types
 of the neighbors
$n$ and $2$; a soft gluon gives rise to the same factorization term
when both neighbors are gluons as it does when one or both neighbors
are quarks.  However, it does depend on whether the soft
gluon is attached to a gluonic part of the loop or to a fermionic part
of the loop.  In \eqn{Anoneformula} all gluons in the leading in
colors contribution are attached to the gluonic part of the loops,
while in the sub-leading in color contributions they are attached to
the fermionic (or scalar) parts of loops.  Since there is no $n_{\!
f}$, $n_s$ or $N_c$ dependence in \eqn{SoftGluonExact} there are no
sub-leading contributions.  This property may also be expressed in
terms of the primitive amplitudes, as we do below.

\subsubsection{Soft factorization of all-gluon primitive amplitudes}
We see from \eqn{SoftGluonExact} that there are no $n_f$ or $n_s$
dependent contributions to soft factorization.  We therefore have the
simple result that
\begin{equation}
\eqalign{
\Soft^{ng,\,[1]}(n,1^\pm,2) =\ & -{c_\Gamma\over\eps^2}
 \Soft^{\tree}(n,1^\pm,2)\left({\mu^2(-s_{n2})\over (-s_{n1})
 (-s_{12})}\right)^{\epsilon}\, {\pi\epsilon\over
 \sin(\pi\epsilon)}\,,\cr 
\Soft^{ng,\,[1/2]}(n,1^\pm,2) =\ & 0\,,\cr
\Soft^{ng,\,[0]}(n,1^\pm,2) =\ & 0\,.\cr
}
\label{PrimSoftAllGlue}
\end{equation}

\subsubsection{Soft factorization of two-quark primitive amplitudes}
As in the case of primitive splitting relations, we must distinguish
between situations where the soft gluon falls between the quark and
the anti-quark in the ordering and those where it falls between the
anti-quark and the quark.  Again casting all factorization properties
in terms of the `$L$'-type primitive amplitudes, we find that when the soft
gluon is between the quark and the anti-quark ({\it i.e.\/} the soft 
gluon is attached to a gluonic part of the loop), the one loop soft
amplitudes are
\begin{equation}
\eqalign{
\Soft^{2q,\,L,\,[1]}(n,1^\pm,2) =\ & -{c_\Gamma\over\eps^2}
 \Soft^{\tree}(n,1^\pm,2)\left({\mu^2(-s_{n2})\over (-s_{n1})
 (-s_{12})}\right)^{\epsilon}\, {\pi\epsilon\over
 \sin(\pi\epsilon)}\,,\cr 
\Soft^{2q,\,L,\,[1/2]}(n,1^\pm,2) =\ & 0\,,\cr
\Soft^{2q,\,L,\,[0]}(n,1^\pm,2) =\ & 0\,.\cr
}
\label{PrimSoftTwoQuarka}
\end{equation}
When the soft gluon is between the anti-quark and the quark so that
it attaches to a fermionic part of the loop, the
one-loop soft amplitudes vanish,
\begin{equation}
\eqalign{
\Soft^{2q,\,L,\,[1]}(n,1^\pm,2) =\ & 0\,,\hskip3in\cr 
\Soft^{2q,\,L,\,[1/2]}(n,1^\pm,2) =\ & 0\,,\cr
\Soft^{2q,\,L,\,[0]}(n,1^\pm,2) =\ & 0\,.\cr
}
\label{PrimSoftTwoQuarkb}
\end{equation}

The case of four or more quarks is similar: if soft gluon in any 
primitive amplitude connects to a fermionic part of the loop
then the soft factor vanishes.  If the soft gluon is attached to a gluonic
part of the loop then the soft factor is given by the universal
function in \eqn{PrimSoftTwoQuarka}.

\subsubsection{Example: The soft limit of
$A_{5;3}(1_{\overline{q}},2_q;4,5;3)$.} 
\label{SoftExampleSubSubSection}

To illustrate the utility of the primitive decomposition of the
one-loop soft and collinear splitting amplitudes, we compute the soft
limit of one of the sub-leading partial amplitudes for two-quark
three-gluon scattering.  Using \eqn{AfivethreeA}, and taking the limit
that gluon $3$ becomes soft,
\begin{equation}
\eqalign{
A_{5;3} & (1_\qb, 2_q ; 4,5; 3) \cr
   \mathop{\longrightarrow}^{k_3 \to 0} \null & \null
  - \left[A_4^L(1_\qb, 4, 5, 2_q) + A_4^L(1_\qb, 5, 4, 2_q)\right]
  \Soft^\tree(1,2)\cr
 &+ A_4^L(1_\qb, 2_q, 4, 5)\left[\Soft^\tree(2,4)
             + \Soft^\tree(4,5) + \Soft^\tree(5,1)\right]\cr
 &+ A_4^L(1_\qb, 2_q, 5, 4)\left[\Soft^\tree(2,5)
             + \Soft^\tree(5,4) + \Soft^\tree(4,1)\right]\cr
 &+ A_4^L(1_\qb, 4, 2_q, 5)\left[\Soft^\tree(2,5)
             + \Soft^\tree(5,1) \right] \cr
 &+ A_4^L(1_\qb, 5, 2_q, 4) \left[ \Soft^\tree(4,1)
             + \Soft^\tree(2,4)\right]\cr
 &- A_4^\tree(1_\qb, 2_q, 4, 5)\left[\Soft^\oneloop(1,2)
             + \Soft^\oneloop(2,5) + \Soft^\oneloop(5,4)
             + \Soft^\oneloop(4,1)\right]\cr
 &- A_4^\tree(1_\qb, 2_q, 5, 4)\left[\Soft^\oneloop(1,2)
             + \Soft^\oneloop(2,4) + \Soft^\oneloop(4,5)
             + \Soft^\oneloop(5,1)\right]\,,\cr
}
\label{A53softlim}
\end{equation}
where we abbreviate $\Soft^\tree(i,j)\equiv\Soft^\tree(i,3,j)$ and
$\Soft^\oneloop(i,j)\equiv\Soft^\oneloop(i,3,j)$.
Some useful properties for taking the limit are $\Soft^\tree(i,3,j) = 
 - \Soft^\tree(j,3,i)$ and
\begin{equation}
A_4^\tree(1_\qb, 4, 2_q, 5) =  A_4^\tree(1_\qb, 5, 2_q, 4) = 
- A_4^\tree(1_\qb, 2_q, 4, 5) - A_4^\tree(1_\qb, 2_q, 5, 4) \,,
\end{equation}
As mentioned in \sec{sec:ColorRev} the tree-level primitive amplitudes
correspond to the partial amplitudes when all particles including the
fermions are taken to be in the adjoint representation.  These
properties of the tree-level primitive amplitudes are simply the
properties~\cite{mpReview} of any such partial amplitudes.

Comparing this result to \eqns{Afourthree}{AfivethreeA}, it is
clear that there is no simple soft factorization relation for
sub-leading color partial amplitudes.  In particular, there 
is no direct analog to
\eqn{OneLoopSoft} which would express the factorization in terms of
lower point partial amplitudes.  The reason for this is that soft
factorization depends upon the soft gluon's neighbors in the color
flow.  Partial amplitudes, especially sub-leading partial amplitudes,
can have a rather complicated color flow~\cite{bdk3g2q} with the
result that soft factorization becomes deeply entangled.  Primitive
amplitudes have a fixed ordering of external legs so there can be no
twist in the color flow to complicate soft factorization. 
By piecing together these primitive components, the soft limits of
sub-leading partial amplitudes can be systematically understood.

\section{Checks on results}
\label{sec:Check}

\subsection{Checks using supersymmetry.}

Although QCD is not a supersymmetric theory we may use supersymmetry
as a way of verifying our results.  As described in the second
appendix of ref.~\cite{bddkSusy4}, we may convert the QCD splitting
functions to those for $N=1$ super-Yang-Mills theory, which contains a
single gluon and gluino, by adjusting the color factors to be $1/
N_c^2 \to -1$, $n_{\! f}/N_c \to 1$ and $n_s/N_c \to 0$ in the
splitting amplitudes (eqs. (\ref{fullggga}) - (\ref{fullqqg})). 
(As discussed in appendix~B of ref.~\cite{bddkSusy4}, one can also 
check the case of an $N=1$ chiral multiplet, but then one must
account for Yukawa interactions.)  To
preserve the supersymmetry we take the parameter controlling
the variant of dimensional regularization to be $\delta_R = 0$.

With these substitutions, the only non-vanishing splitting amplitudes
are those where the tree-level amplitudes do not vanish.  All the
remaining non-vanishing splitting amplitudes are given by
\begin{equation}
\Split_{-\lambda}(1^{\lambda_1},
  2^{\lambda_2}) = c_\Gamma
  \Split^{\tree}_{-\lambda}(1^{\lambda_1},2^{\lambda_2}) \,
  r_S(-\lambda,1^{\lambda_1},2^{\lambda_2}) \,,
\label{SusySplit}
\end{equation}
where 
\begin{equation}
 r_S(-\lambda,1^{\lambda_1},2^{\lambda_2}) = f_2(s_{12}, \eps) 
+ f_3(s_{12}, \eps, z) + f_3(s_{12}, \eps, 1-z) \,,
\label{SusyRS}
\end{equation}
is independent of the specific helicities or particle labels.  (The
explicit value of the sum of these functions is given in
\eqn{nonfactall}.)  This independence is a consequence of
supersymmetry~\cite{SWI} as has been previously noted through
$\Ord(\eps^0)$~\cite{bddkSusy4}.  The fact that our expressions (for
$\delta_R = 0$) satisfy this constraint to all orders in $\eps$
provides non-trivial support that the results are correct.

For the special case of $N=4$ super-Yang-Mills explicit results 
for the one-loop four- five- and six-point amplitudes are known 
to all orders in $\eps$~\cite{DimShift}.  Using these results, we may extract
the splitting amplitudes simply by taking the collinear limit of the
five-point amplitude and comparing it to the four-point amplitudes.
The result so obtained agrees with the supersymmetric results
in \eqns{SusySplit}{SusyRS}, providing another independent check.

We have also performed the same checks in the case where an external
gluon momentum becomes soft.  In this case the supersymmetry is manifest
since the soft amplitude does not depend on the number of flavors
or colors or on whether the neighboring particles are gluons or fermions.

\subsection{Checks from Higgs amplitudes}

We have also checked the one-loop $g\rightarrow gg$ and $g\rightarrow
q\qb$  splitting amplitudes, as well as the one-loop soft amplitude,
by directly computing to all orders in $\eps$ the one-loop $gggH$ and
$gq\qb{H}$ amplitudes in a theory with an effective $ggH$
coupling~\cite{Dawson}
\begin{equation}
{\cal L}_{\rm eff}\ =\ -{1\over4}\left[1-{\alpha_{s}\over3\pi}{H\over 
v}\left(1+{11\alpha_{s}\over4\pi}\right)\right]\,
{\rm Tr}\,G_{\mu\nu}G^{\mu\nu}\, ,
\label{Hggcoupling}
\end{equation}
given by the infinite mass limit of a heavy quark triangle.  The
$gggH$ and $gq\bar qH$ amplitudes in this theory are convenient for
extracting the splitting and soft amplitudes since they involve
only four-point kinematics with a single massive leg; the massive
Higgs ensures that both amplitudes have well-defined soft and
collinear limits.  Since there is no tree-level $q\qb{H}$ coupling in
this effective theory, we cannot use these processes, however, to check $q\to
qg$ splitting.

The details of the calculation through ${\cal O}(\eps^0)$ may be 
found in ref.~\cite{crs}.  For this comparison we have extended the
calculation to all orders in $\eps$.  We find that the amplitudes can
be written as simple analytic expressions with the scalar box
integrals expressed in terms of hypergeometric functions, as given in
\app{app:BoxApp}.

Since there are only three external partons, there are no sub-leading
partial amplitudes for these processes.  Primitive decompositions are
trivial to perform.  The $gggH$ amplitudes mirror the all-gluon
amplitudes of pure QCD with the primitive amplitudes given simply by 
the parts of
the leading partial amplitude corresponding to the different
spins of the loop particle.  The $gq\qb{H}$ amplitudes mirror the
two-quark amplitudes of pure QCD, with the rules for those orderings
with the gluon between the quark and anti-quark $(1_\qb,2_q,3)$
differing from the rules for those with the gluon between the
anti-quark and quark $(1_\qb,3,2_q)$.

With the $gggH$ amplitudes, we have checked the one-loop $g\to gg$
splitting amplitudes (\eqns{fullggga}{fullgggb}) as well as the
one-loop soft amplitudes (\eqn{SoftGluonExact}).  With the $gq\qb{H}$
amplitudes, we can check the one-loop $g\to\qb{q}$ splitting
amplitudes (eqs.~(\ref{fullgqq}), (\ref{qqPrSplgqqa}) and
(\ref{qqPrSplgqqb})).  In all cases, we find complete agreement to all
orders in $\eps$.  These checks emphasize the universal, {\it i.e.\/}
process independent, nature of infrared factorization.

\section{Conclusions}

In this paper we have presented the functions describing the behavior
of one-loop amplitudes in the infrared divergent regions of phase
space.  These functions should be useful for NNLO $n$-jet
computations.  When evaluating the phase space integrals associated
with the $(n+1)$-point one-loop amplitudes with one unresolved parton,
infrared divergences of order $\eps^{-2}$ in the dimensional
regularization parameter are encountered.  This suggests that one
needs to know the one-loop amplitudes through $\Ord(\eps^2)$.  Given
the complex analytic structure of multi-parton
amplitudes~\cite{bdkZ4,cgmZ4,bdk5g,kst1g4q,bdk3g2q,dksWW}, it would be
rather non-trivial to obtain the complete higher order in $\eps$ 
contributions.  A better approach would be to use the splitting functions 
presented in this paper to all orders in $\eps$ to obtain the higher order 
in $\eps$ contributions only in the relevant soft and collinear regions of 
phase space.  Then the only calculations needed would be the 
$(n+1)$-point one-loop amplitudes through $\Ord(\eps^0)$ and the 
$n$-point amplitudes through
$\Ord(\eps^2)$, which is much simpler to obtain.  We also
discussed the decomposition of QCD amplitudes into primitive amplitudes
as a way to disentangle the color factors from the infrared divergent
regions of phase space; for the case of external fermions these
primitive amplitudes are a finer subdivision of the amplitudes than
the conventional color decomposed ones.

In order to extend the analysis in this paper to the case of two-loop
soft and collinear splitting amplitudes, one would need a detailed
understanding of the infrared structure of two-loop integrals as well
as of the discontinuity functions which may arise in the soft or
collinear limits.  The factorizing contributions are, however,
straightforward to compute since they involve only triangle
contributions.

The development of formalisms that would enable one to compute the
NNLO contributions to multi-jet amplitudes remains an important open
problem in perturbative QCD.  There has, however, been some recent
progress towards this.  For example, the singularity structure when
two partons are unresolved has been examined in ref.~\cite{cg}.  An
interesting formalism for obtaining collinear splitting amplitudes
from unitarity~\cite{bdkReview,bddkSusy4,bddkSusy1} has also recently
been developed~\cite{KosowerSplit}. Furthermore, two-loop four-gluon
amplitudes in the special case of maximal supersymmetry have been
evaluated in terms of scalar integral functions~\cite{bry} using the
new techniques described in ref.~\cite{bdkReview}.  Although a number
of substantial difficulties remain, these developments provide for
some optimism.

\vskip .2 cm

\section*{Acknowledgments}
 
We wish to thank H. Braden and A.O. Daalhuis for discussions on
hypergeometric functions and D.A. Kosower for discussions on splitting
functions.  We also thank D.A. Kosower and P. Uwer for showing us a
draft of their paper on one-loop splitting amplitudes to all orders in the
dimensional regularization parameter~\cite{KosowerFuture}. Their results
are obtained using the new method for computing splitting amplitudes 
from unitarity described in ref.~\cite{KosowerSplit}.

The work of Z.B. was supported by the US Department of Energy under
grant DE-FG03-91ER40662, that of W.B.K. by the US Department of Energy
under grant DE-AC02-98CH10886 and that of C.R.S. by the US National
Science Foundation under grant PHY-9722144. The work of V.D.D. and
C.R.S was also supported by NATO Collaborative Research Grant
CRG-950176.


\appendix

\section{Tree-level splitting amplitudes}
\label{app:TreeSplitAppendix}

We collect here the helicity form of the tree-level
splitting amplitudes~\cite{mpReview}, using the sign conventions
of ref.~\cite{bdk3g2q}. For the case of a gluon
splitting into two gluons, with helicities labeled as if all particles
are outgoing, the splitting amplitudes are
\begin{equation}
\eqalign{
\eqalign{
{\Split}^{g\to g_1g_2,\,{\tree}}_{+}(1^{+},2^{+})
             &= 0 \,,\cr
{\Split}^{g\to g_1g_2,\,{\tree}}_{+}(1^{-},2^{+})
             &= {z^2\over \sqrt{z (1-z)} \spa1.2 } \,,\cr}
\hskip 10pt\eqalign{
{\Split}^{g\to g_1g_2,\,{\tree}}_{+}(1^{-},2^{-})
             &= {-1\over \sqrt{z (1-z)} \spb1.2 } \,,  \cr
{\Split}^{g\to g_1g_2,\,{\tree}}_{+}(1^{+},2^{-})
             &= {(1-z)^2\over \sqrt{z (1-z)} \spa1.2 } \,, \cr}\cr
\eqalign{
{\Split}^{g\to g_1g_2,\,{\tree}}_{-}(1^{-},2^{-})
             &= 0 \,,\cr
{\Split}^{g\to g_1g_2,\,{\tree}}_{-}(1^{+},2^{-})
             &= {-z^2\over \sqrt{z (1-z)} \spb1.2 } \,,\cr}
\hskip 10pt\eqalign{
{\Split}^{g\to g_1g_2,\,{\tree}}_{-}(1^{+},2^{+})
             &= {1\over \sqrt{z (1-z)} \spa1.2 } \,,  \cr
{\Split}^{g\to g_1g_2,\,{\tree}}_{-}(1^{-},2^{+})
             &= {-(1-z)^2\over \sqrt{z (1-z)} \spb1.2 } \,. \cr}
}
\label{gggTreeSplit}
\end{equation}
The spinor inner products~\cite{mpReview,SpinorHelicity} are
$\spa{i}.j = \langle i^- | j^+\rangle$ and $\spb{i}.j = \langle i^+|
j^-\rangle$, where $|i^{\pm}\rangle$ are massless Weyl spinors of
momentum $k_i$, labeled with the sign of the helicity.  They are
antisymmetric, with norm $|\spa{i}.j| = |\spb{i}.j| = \sqrt{s_{ij}}$,
and obey the rule $\spa{i}.j\spb{j}.i = s_{ij} = 2k_i\cdot k_j$. 

For $g\to\qb_1q_2$,
\begin{equation}
\eqalign{
\Split^{g\to\qb_1q_2,\,{\tree}}_{+}(1_\qb^{+},2_q^{-})
           &= {z^{1/2}(1-z)^{3/2}
          \over \sqrt{z(1-z)}\spa{1}.{2}}\,,\cr
\Split^{g\to\qb_1q_2,\,{\tree}}_{-}(1_\qb^{-},2_q^{+})
           &= {z^{1/2}(1-z)^{3/2}
          \over \sqrt{z(1-z)}\spb{1}.{2}}\,,\cr}
\hskip 10pt\eqalign{
\Split^{g\to\qb_1q_2,\,{\tree}}_{+}(1_\qb^{-},2_q^{+})
           &= {z^{3/2}(1-z)^{1/2}
          \over \sqrt{z(1-z)}\spa{1}.{2}}\,,\cr
\Split^{g\to\qb_1q_2,\,{\tree}}_{-}(1_\qb^{+},2_q^{-})
           &= {z^{3/2}(1-z)^{1/2}
          \over \sqrt{z(1-z)}\spb{1}.{2}}\,.\cr}
\label{gqqTreeSplita}
\end{equation}

Finally, for $q\to q_1g_2$ and $\qb\to g_1\qb_2$,
\begin{equation}
\eqalign{
\Split^{q\to q_1g_2,\,{\tree}}_{-}(1_q^{+},2^{+})
          &= {z^{1/2}\over \sqrt{z(1-z)}\spa{1}.2}\,,\cr
\Split^{q\to q_1g_2,\,{\tree}}_{+}(1_q^{-},2^{-})
          &= {-z^{1/2}\over \sqrt{z(1-z)}\spb{1}.2}\,,\cr
\Split^{\qb\to g_1\qb_2,\,{\tree}}_{-}(1^{+},2_\qb^{+})
          &= {(1-z)^{1/2}\over \sqrt{z(1-z)}\spa{1}.{2}}\,,\cr
\Split^{\qb\to g_1\qb_2,\,{\tree}}_{+}(1^{-},2_\qb^{-})
          &= {-(1-z)^{1/2}\over \sqrt{z(1-z)}\spb{1}.{2}}\,,\cr}
\hskip10pt\eqalign{
\Split^{q\to q_1g_2,\,{\tree}}_{-}(1_q^{+},2^{-})
          &= {-z^{3/2}\over \sqrt{z(1-z)}\spb{1}.2}\,, \cr
\Split^{q\to q_1g_2,\,{\tree}}_{+}(1_q^{-},2^{+})
          &= {z^{3/2}\over \sqrt{z(1-z)}\spa{1}.2}\,, \cr
\Split^{\qb\to g_1\qb_2,\,{\tree}}_{-}(1^{-},2_\qb^{+})
          &= {-(1-z)^{3/2}\over \sqrt{z(1-z)}\spb{1}.{2}}\,,\cr
\Split^{\qb\to g_1\qb_2,\,{\tree}}_{+}(1^{+},2_\qb^{-})
          &= {(1-z)^{3/2}\over \sqrt{z(1-z)}\spa{1}.{2}}\,.\cr}
\label{qqgTreeSplita}
\end{equation}

Because we are working with color ordered amplitudes, we must
distinguish between, say, $q\to q_1g_2$ and $q\to g_1q_2$.  This is
accomplished by exchanging $1$ and $2$ (and therefore $z$ and $1-z$).
For example,
\begin{equation}
\Split^{g\to q_1\qb_2,\,{\tree}}_{+}(1_q^{-},2_\qb^{+}) =
{z^{3/2}(1-z)^{1/2} \over \sqrt{z(1-z)}\spa{2}.{1}} =
{-z^{3/2}(1-z)^{1/2} \over \sqrt{z(1-z)}\spa{1}.{2}}\,.
\label{splitreorder}
\end{equation}

\section{Previous results through $O(\eps^0)$ for one-loop 
splitting amplitudes.}
\label{app:LoopSplitAppendix}

For convenience we collect the splitting and soft amplitudes through
$\Ord(\eps^0)$. These were obtained from refs.~\cite{bddkSusy4,bdk3g2q}.
These are useful since they contain the infrared singularities
which are used to fix the non-factorizing contributions.  They 
also provides a check on our results.
 
The loop splitting amplitudes have a structure similar to the
tree splitting amplitudes, so it is useful to express them in terms
of a proportionality constant $r_S$ defined by
\begin{equation}
 \Split^{\oneloop}_{-\lambda}(1^{\lambda_1},2^{\lambda_2})
 = c_\Gamma
 \times \Split^{\tree}_{-\lambda}(1^{\lambda_1},2^{\lambda_2})
 \times r_S(-\lambda,1^{\lambda_1},2^{\lambda_2}) \,,
\label{genrsdef}
\end{equation}
for general partons $1$ and $2$.
The only exception to \eqn{genrsdef} is for $g^-\to g^+g^+$
(and its parity conjugate $g^+\to g^-g^-$), where $\Split^{\tree}$
vanishes but $\Split^{\oneloop}$ does not.
In general $r_S(-\lambda,1^{\lambda_1},2^{\lambda_2})$ depends
on the parton helicities.
As a notational point we distinguish between particles circulating
in the loop by a spin index, {\it i.e.\/} $\Split^{[J]}(1,2)$.  For the cases
$J=0,1/2,1$ respectively correspond to a scalar, a fermion, and a gluon
in the loop.

The $\Split^{[J]}_{+}(1^{+},2^{+})$ obey the supersymmetry relation
$\Split^{[1]} = -\Split^{[1/2]} = \Split^{[0]}$, where
\begin{equation}
\Split^{[1]}_{+}(1^{+},2^{+})\ =\
-{1\over 48\pi^2}
  \sqrt{z(1-z)}\ {\spb{1}.2  \over {\spa{1}.2}^2}\ .
\label{finitegggloop}
\end{equation}
We present the remaining $g\to gg$ loop splitting amplitudes
in terms of $r_S$:
\begin{equation}
\eqalign{
r^{g\to g_1g_2,\,{\oneloop}}_{S}(\pm,1^\mp,2^\mp) & =
 - {1\over\eps^2}\L{\mu^2\over z(1-z)(-s_{12})}\R^{\eps}
 + 2 \ln z\,\ln(1-z) - {\pi^2\over6} \cr
& \hskip 3 cm 
 + {1\over 3} z(1-z) \left(1-\eps\delta_R - {n_f\over N_c}
 +{n_s\over N_c}\right)
\, , \cr
r^{g\to g_1g_2,\,{\oneloop}}_{S}(\lambda,1^\pm,2^\mp) & = 
 - {1\over\eps^2}\L{\mu^2\over z(1-z)(-s_{12})}\R^{\eps}
 + 2 \ln z\,\ln(1-z) - {\pi^2\over6} \,. \cr}
\label{gggrenorm}
\end{equation}

The $r_S$ functions for the case 
\begin{equation}\eqalign{
r^{q\to q_1g_2,\,{\oneloop}}_S(\pm,1^{\mp},2^{\mp})
  & =  f(1-z,s_{12}) + \left(1+{1\over N_c^2}\right) {1-z\over2}\, ,\cr
r^{q\to q_1g_2,\,{\oneloop}}_S(\pm,1^{\mp},2^\pm) 
&   = f(1-z,s_{12})\ ,\cr
r^{\qb\to g_1\qb_2,\,{\oneloop}}_S(\pm,1^{\mp},2^{\mp})
 &  = f(z,s_{12}) + \left(1+{1\over N_c^2}\right) {z\over2}\, ,\cr
r^{\qb\to g_1\qb_2,\,{\oneloop}}_S(\pm,1^{\mp},2^\pm) 
 &  = f(z,s_{12})\, ,\cr}
\label{qgqrenorm}
\end{equation}
where the function $f(z,s)$ is
\begin{equation}
\eqalign{
  f(z,s)\ &=\ - {1\over\eps^2}\L{\mu^2\over z(-s)}\R^{\eps} - \Li_2(1-z) \cr
  &\quad  -{1\over N_c^2} \left[
     - {1\over\eps^2}\L{\mu^2\over(1-z)(-s)}\R^{\eps}
     + {1\over\eps^2}\L{\mu^2\over(-s)}\R^{\eps} - \Li_2(z) \right]\,. \cr}
\label{fdef}
\end{equation}

For $g\to\qb_1q_2$ the results are
\begin{equation}\eqalign{
r^{g\to\qb_1q_2,\,{\oneloop}}_{S}(\lambda,1^\pm,2^\mp) &=\
  -{1\over\eps^2} \left[ \L{\mu^2\over z(-s_{12})}\R^{\eps}
                  + \L{\mu^2\over (1-z)(-s_{12})}\R^{\eps}
                  - 2\L{\mu^2\over -s_{12}}\R^{\eps} \right] \cr
&\qquad
  + {13\over6\eps} \L{\mu^2\over (-s_{12})}\R^{\eps}
  + \ln(z)\,\ln(1-z) - {\pi^2\over6} + {83\over18}
                                     - {\delta_R\over6} \cr
&\qquad   -{1\over N_c^2} \left[
    -{1\over\eps^2} \L{\mu^2\over -s_{12}}\R^{\eps}
    -{3\over2\eps} \L{\mu^2\over -s_{12}}\R^{\eps} - {7\over2}
                                     - {\delta_R\over2} \right]\, \cr
 &\qquad   - {n_{\! f}\over N_c}\left[{2 \over3\eps}
      \L{\mu^2 \over -s_{12}}\R^\eps + {10\over9}\right]
   -{n_s\over N_c}\left[{1\over3\eps} \L{\mu^2 \over -s_{12}}\R^\eps +
    {8\over9}\right]\, .\cr}
\label{qqgrenorm}
\end{equation}


\section{Evaluation of box integrals in collinear and soft limits}
\label{app:BoxApp}

In this appendix we evaluate the integral functions appearing
in the one-loop soft and collinear splitting amplitudes
to all order in $\eps$.  In the collinear or soft limits, the hypergeometric
functions in terms of which the integrals may be expressed reduce
to sums of polylogarithms.  These results allow us to present
explicit formulas for the non-factorizing contributions, 
valid to all orders in dimensional regularization.

%
\begin{figure}[ht]
\centerline{\epsfxsize 1.2 truein \epsfbox{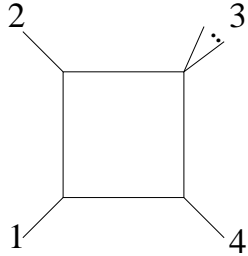}}
\vskip -.2 cm
\caption[]{
\label{BoxFigure}
\small The particular box integral under consideration in this
appendix. Leg 3 is composed of a sum over massless momenta so that
$K_3^2 \not = 0$.}
\end{figure}

Consider the single-mass box integral ${\cal I}_{4:3}^{1m}(s,t, K_3^2)$,
depicted in \fig{BoxFigure}, 
where we take the third leg to be massive, $K_3^2 \not = 0$
 and
the Mandelstam variables appearing as arguments are,
\begin{equation}
s = (k_1+ k_2)^2\, \hskip 2 cm 
t = (k_1+ k_4)^2  \,.
\label{invar}
\end{equation}
Limits of this integral appear in the collinear functions $f_3$ in
\eqn{ffunctions} and in the soft function in \eqn{soft}.
All other box integrals appearing in the soft and collinear 
splitting amplitudes
are given by relabelings of this integral.

It can be expressed as a sum of hypergeometric functions~\cite{IntegralLong}
\begin{equation}
\eqalign{
  - i \mu^{2\eps} st {\cal I}^{\rm 1m}_{4:3} & = 
  {2c_\Gamma \mu^{2\eps}\over \eps^2}\left\{
    \left({t-K_3^2\over st}\right)^{\eps}
    \hypgeo\!\!\left(-\eps,-\eps;1-\eps;1+{s\over t-K_3^2}\right)\right.\cr
  &\phantom{{2r_\Gamma\over st}}+\left({s-K_3^2\over st}\right)^{\eps}
    \hypgeo\!\!\left(-\eps,-\eps;1-\eps;1+{t\over s-K_3^2}\right)\cr
  &\left.\phantom{{2r_\Gamma\over st}}
    -\left({(s-K_3^2)(t-K_3^2)\over -stK_3^2}\right)^{\eps}
    \hypgeo\!\!\left(-\eps,-\eps;1-\eps;1-{st\over(s-K_3^2)
    (t-K_3^2)}\right)\right\}\,. \cr}
\label{exact}
\end{equation}

\subsection{The collinear limit of the single-mass box integral}
\label{app:ColBox}
In the collinear limit $k_1=zP$ and $k_2=(1-z)P$, we find
that $s\to 0$, $t\to z K_3^2$, and \eqn{exact}
reduces to
\begin{equation}
- i \mu^{2\eps} st {\cal I}_{4:3}^{\rm 1m}(s,t,K_3^2) \iscol{k_1}{k_2}\
{2c_{\Gamma}\over\epsilon^2}\, 
\left({\mu^2\over -s}\right)^{\epsilon}\, z^{-\epsilon}\,
 \hypgeo \left(-\epsilon, -\epsilon; 1-\epsilon; 1-z\right)\,.
\label{coll}
\end{equation}
We wish to re-express this as a power series in $\eps$.
Using the hypergeometric identity
\begin{equation}
\hypgeo \left(-\epsilon, -\epsilon; 1-\epsilon; 1-z\right) =
z^{\epsilon}\, \hypgeo \left(1, -\epsilon; 1-\epsilon; 
 -{1-z\over z}\right)\, ,
\end{equation}
and the expansion of the hypergeometric function as a power series in $z$,
\begin{equation}
\eqalign{
  \hypgeo(1,-\eps;1-\eps;z) & = 1+\sum_{n=1}^{\infty}
    {(n!)(-\eps)(1-\eps)\cdots(n-1-\eps)\over(1-\eps)(2-\eps)
    \cdots(n-\eps)(n!)}z^n\cr
 & = 1-\sum_{m=1}^{\infty}\eps^m\left(\sum_{n=1}^\infty
    {z^n\over n^m}\right)= 1-\sum_{m=1}^{\infty}\eps^m
    {\Li}_m(z)\,,\cr}
\end{equation}
we can write the single external mass box integral (\eqn{coll}) as
\begin{equation}
\eqalign{
- i \mu^{2\eps} st {\cal I}_{4:3}^{\rm 1m}(s,t,K_3^2)
& \iscol{k_1}{k_2} {2c_\Gamma\over\eps^2}
    \left({\mu^2\over-s}\right)^{\eps}
    \hypgeo \left(1,-\eps;1-\eps;-{1-z\over z}\right)\cr
  &  = {2c_\Gamma\over\eps^2}\left({\mu^2\over-s}\right)^{\eps}
    \left(1-\sum_{m=1}^{\infty}\eps^m
    {\Li}_m\left(-{1-z\over z}\right)\right) \,.\cr}
\label{massbox}
\end{equation}
Note that this contains the infrared singularity,
\begin{equation}
{{2c_\Gamma\over\eps}{\Li}_1\left(-{(1-z)\over z}\right)
  =  {2c_\Gamma\over\eps}\ln(z)\,.}
\end{equation}
As discussed in \sec{sec:LoopSofCol} the known coefficient of this
singularity may be used to determine the overall coefficient of the
function $f_3(s_{12}, \eps, z)$ given in \eqn{ffunctions}. Similarly, the
coefficient of $f_3(s_{12}, \eps, 1-z)$ is fixed from the coefficient
of ${2c_\Gamma \ln(1-z) / \eps}$.

When the splitting amplitude contains the combination $f_3(s, \eps, z)
+ f_3(s, \eps, 1-z)$, as in the pure gluon case, we may further
simplify the expression.  Using hypergeometric identities (see
eq.~(9.132) of ref.~\cite{GradRyz}), we find that
\begin{equation}
\eqalign{
  \hypgeo \left(1,-\eps;1-\eps;z^{-1}\right) 
& = - z\, {\Gamma(1-\eps) \Gamma(-1-\eps) \over \Gamma^2(-\eps)}\, 
        \hypgeo(1,1+\eps, 2+\eps, z)  \cr
& \hskip 3 cm 
    + (-z)^{-\eps} \Gamma(1-\eps) \Gamma(1+\eps) \, 
         \hypgeo(-\eps, 0, -\eps, z) \cr
& = -z\, {\eps\over1+\eps}\, \hypgeo\!\!\left(1,1+\eps;2+\eps;z\right)
  + (-z)^{-\eps} \Gamma(1-\eps)\Gamma(1+\eps)\cr
& = 1-\hypgeo\left(1,\eps;1+\eps;z\right)
  +(-z)^{-\eps}\Gamma(1-\eps)\Gamma(1+\eps)\,,\cr}
\end{equation}
so that we can rewrite $- i \mu^{2\eps} st {\cal I}^{1m}_{4;3}(s,t)$ as
\begin{equation}
- i \mu^{2\eps} st {\cal I}^{\rm 1m}_{4;3}(s,t)=
  {2c_\Gamma\over\eps^2}\left({\mu^2\over-s}\right)^{\eps}
  \left[1-\hypgeo\left(1,\eps;1+\eps; {-z\over1-z}\right)
  +\left({1-z\over z}\right)^{\eps}\Gamma(1-\eps)
  \Gamma(1+\eps)\right] \,.
\end{equation}

Thus, using \eqn{ffunctions}, the total non-factorizing
contribution for the case of $g\to gg$ splitting, given in
\eqn{rsgluons} is
\begin{equation}
\eqalign{
  r_S & = f_2(s, \epsilon) + f_3(s,\epsilon,z) 
     + f_3(s,\epsilon, 1-z)\cr
& ={1\over\eps^2}\left({\mu^2\over-s}\right)^{\eps}
    \left[1-\hypgeo \left(1,-\eps;1-\eps;
    -{z\over1-z}\right) - \hypgeo\left(1,-\eps;1-\eps;
    {z-1\over z}\right)\right]\cr
&  = {1\over\eps^2}\left({\mu^2\over-s}\right)^{\eps}\left[
    2\sum_{m=1,3,5,\dots}^\infty\eps^m\left(\sum_{n=1}^\infty
    {1\over n^m}\left({-z\over1-z}\right)^n\right)
    -\left({1-z\over z}\right)^{\eps}
\Gamma(1-\epsilon)\Gamma(1+\epsilon)\right]\cr
& = {1\over\eps^2}\left({\mu^2\over-s}\right)^{\eps}
\left[ 2 \sum_{m=1,3,5,...}^\infty \epsilon^m 
{\Li}_m\left({-z\over 1-z}\right) 
-\left({1-z\over z}\right)^\epsilon 
{\pi\epsilon\over \sin(\pi\epsilon)} \right]\,. \cr}
\label{nonfactall}
\end{equation}

\subsection{The soft limit of the single-mass box integral}
\label{app:SoftFun}
In the soft limit $k_1\to 0$, we have $s\to 0$, $t\to 0$, and
the single mass box integral \eqn{exact}, reduces to
\begin{equation}
\Biggl[- i \mu^{2\eps} st {\cal I}_{4:3}^{1m}(s,t,K_3^2)\Biggr]_{k_1\to 0} 
= {2c_{\Gamma}\over\epsilon^2}\, 
\left({\mu^2(-K_3^2)\over (-s)(-t)}\right)^{\epsilon}\, 
_{2}F_1\left(-\epsilon, -\epsilon; 1-\epsilon; 1 \right)\,.
\label{softa}
\end{equation}
Using the identity
\begin{equation}
_{2}F_1\left(-\epsilon, -\epsilon; 1-\epsilon; 1\right)
= {\pi\epsilon\over \sin(\pi\epsilon)}\,,
\end{equation}
we can write the soft discontinuity function $f_5(s_{n1}, s_{12},
s_{n2}, \eps)$ to all orders in $\epsilon$ as
\begin{equation}
f_5(s_{n1}, s_{12}, s_{n2}, \eps) = -{c_{\Gamma}\over\epsilon^2}\,
\left({\mu^2(-s_{ab})\over (-s_{as})(-s_{sb})}\right)^{\epsilon}\, 
{\pi\epsilon\over \sin(\pi\epsilon)}\,.
\label{soft}
\end{equation}


\end{document}